\documentclass{svjour3}
\journalname{Journal of Statistical Physics}
\usepackage{amsmath}
\usepackage{amssymb}
\usepackage{multirow}
\usepackage{graphicx}

\newcommand{\ee}{\mathrm{e}}
\newcommand{\dd}{\mathrm{d}}
\newcommand{\ii}{\mathrm i}
\newcommand{\bx}{{\boldsymbol x}}
\newcommand{\by}{{\boldsymbol y}}
\newcommand{\hq}[0]{\hat{Q}_6}
\newcommand{\hc}[0]{\hat{C}}
\newcommand{\q}[0]{Q_6}

\newcommand{\V}[1]{\boldsymbol{#1}}
\newcommand{\mean}[1]{\left\langle {#1} \right\rangle }
\newcommand{\paren}[1]{\left( {#1} \right ) }

\newcommand{\caja}[1]{\left[ {#1} \right ] }
\newcommand{\grad}[0]{\boldsymbol{\nabla}}
\newcommand{\Ms}{M_{\mathrm s}}

\begin{document}

\title{Tethered Monte Carlo: Managing rugged free-energy landscapes with a Helmholtz-potential formalism}

\author{V.~Martin-Mayor\and B.~Seoane\and D.~Yllanes}

\institute{
V.~Martin-Mayor \and B.~Seoane\and D.~Yllanes\at
Instituto de Biocomputaci\'on y F\'{\i}sica de
Sistemas Complejos (BIFI), Zaragoza, Spain\\
Departamento de F\'\i{}sica Te\'orica I, Universidad
  Complutense, 28040 Madrid, Spain.
}
\maketitle

\begin{abstract}
Tethering methods allow us to perform Monte Carlo simulations in ensembles
with conserved quantities. Specifically, one couples a reservoir to the
physical magnitude of interest, and studies the statistical ensemble where the
total magnitude (system+reservoir) is conserved. The reservoir is actually
integrated out, which leaves us with a fluctuation-dissipation formalism that
allows us to recover the appropriate Helmholtz effective potential with great
accuracy.  These methods are demonstrating a remarkable flexibility. In fact,
we illustrate two very different applications: hard spheres crystallization
and the phase transition of the diluted antiferromagnet in a field (the
physical realization of the random field Ising model). The tethered approach
holds the promise to transform cartoon drawings of corrugated free-energy
landscapes into real computations. Besides, it reduces the algorithmic dynamic
slowing-down, probably because the conservation law holds non-locally.
\end{abstract}

\section{Introduction}
Monte Carlo simulation is one among the handful of general methods that
physicists can use to explore strongly coupled problems, far from the
perturbative regime~\cite{landau:05}. Furthermore, the Monte Carlo method has
become itself an object of active investigations. The reason is twofold: since
it is one of our most cherished tools we wish to sharpen it. But, it is also
true that the dynamic bottlenecks found during the simulation resemble closely
the dynamic arrests that one may find in Nature.

Here, we discuss a strategy to address a fairly general problem in Monte Carlo
simulations: free-energy barriers. Indeed, the time that a standard, canonical
simulation gets trapped in a local minimum grows exponentially
with the free-energy barrier to be surmounted. Let us recall some important
instances of this generic problem (the list is far from exhaustive):
\begin{enumerate}
\item In every first-order transition the system spontaneously segregates in
  a spatially heterogeneous mixture of the two phases (think, e.g., of ice and
  water at the melting point). The two coexisting phases are separated by an
  interface. As a consequence, in order to produce a significant change, the
  system must build an interface of linear dimensions comparable to those of
  the simulation box, $L$. The corresponding free-energy cost scales with the
  system's cross-section $L^D$, where $D$ is the space
  dimension~\cite{lee:90}. As a consequence, $\tau$, the simulation
  characteristic time grows as $\tau\propto\mathrm{exp}[\Sigma L^{D-1}]$,
  where $\Sigma$ is a surface tension. This disaster is named {\em exponential
    dynamic slowing down.\/}

\item Studies of crystallization, the first-order phase transition encountered
  upon cooling or compressing a liquid, are hampered by problems worse than
  exponential dynamic slowdown. Even for such a simple model liquid as
  hard spheres, there is an impressively large number of free-energy minima
  where the simulation may get stuck. For instance, although for simple model
  liquids the equilibrium crystal is face-centered cubic, the system might be
  prone to nucleate a body-centered cubic phase~\cite{tenwolde:95}. Even if a
  crystal of correct symmetry is formed, it may have a large number of defects
  (point defects, such as vacancies, or non-local defects such as
  dislocations). Furthermore, an amorphous solid (a glass) may
  appear~\cite{zaccarelli:09}.

\item Even if the phase transition is continuous, large barriers might be
  present at the critical temperature, as it is the case for the random field
  Ising model~\cite{nattermann:97}, where $\Delta F\propto L^\theta\ll
  L^{D-1}$. Note that, for this problem, both the appropriate order parameter
  and the microscopic ground state configuration~\cite{ogielski:86} are
  known. However, the simulation gets trapped in local minima with escape times
  $\log \tau\sim L^\theta$. An added, major difficulty is the need to
  simulate a huge number of samples to obtain disorder averages truly
  representative of the system's behavior.

\item In spin glasses~\cite{mezard:87} not even the appropriate order parameter
  is known. To detect the phase transition one must use real replicas (clones
  of the system, evolving under the same Hamiltonian but with uncorrelated
  thermal noise). There is a fairly large number of local minima where the
  simulation may trap (at least for finite systems~\cite{marinari:00}). Some
  specific Monte Carlo methods have been devised for this problem, such as the
  exchange Monte Carlo method~\cite{hukushima:96,marinari:98b} (also known as
  parallel tempering). Furthermore, specific hardware has been used to study
  these systems~\cite{ogielski:85,cruz:01,ballesteros:00,jimenez:05,janus:06,janus:08,janus:08b,janus:09,janus:09b,janus:10,janus:10b}.
 Unfortunately,
  exchange Monte Carlo suffers a strong (probably exponential) dynamic slowing
  down below the critical temperature~\cite{janus:10}.
\end{enumerate}

We have some effective and general methods to cope with the first type of
problem, namely first-order phase transitions where the high and the low-temperature
configurations are clearly identified, and easily
findable. Multicanonical~\cite{berg:92} or Wang-Landau~\cite{wang:01}
simulations feature a generalized statistical ensemble. The system performs a
random walk in the energy space, back and forth from the energy of the ordered
phase to that of the disordered phase. At the price of optimizing a number of
parameters proportional to the system size, the free-energy barrier separating
these two phases is easily overcome in small systems. However, the random-walk
strategy can only delay to larger system sizes the advent of exponential
dynamic slow down. In fact, for large enough systems, surface-energy effects
induce geometric transitions in the energy gap between the ordered and the
disordered
phase~\cite{biskup:02,binder:03,macdowell:04,macdowell:06,nussbaumer:06}. The
energy random walk needs a huge amount of simulation time in order to {\em
  tunnel} through the geometric transitions, which results in exponential
dynamic slowing down~\cite{neuhaus:03}.

A different approach stems from Lustig's microcanonical Monte
Carlo~\cite{lustig:98}. Given its microcanonical nature, one may perform
independent simulations at different energies located in the gap between the
ordered and the disordered phase. In this way, one avoids the tunneling
through the geometric transitions, which allows the equilibration of larger
systems. Using a fluctuation-dissipation formalism one can recover the entropy
density, from which a precision study of the phase transition
follows~\cite{martin-mayor:07}.\footnote{It was emphasized by W. Janke that
  the entropy is a more natural thermodynamic potential than the free energy
  for the analysis of first-order transitions~\cite{janke:98}.}

However, in general applications, the {\em gap} between coexisting phases is
not an energy gap. Rather, some reaction coordinate (such as an order
parameter) labels the different regions of the phase space that we wish to
explore. For instance, liquid or crystalline configurations in hard spheres
all have the same energy. Just as entropy is the natural thermodynamic
potential when the reaction coordinate is the internal energy, in the general
case one wishes to study the Helmholtz effective potential associated to the
appropriate reaction coordinate. This is the case, for instance, in the study
of crystallization kinetics in supercooled liquids~\cite{tenwolde:95,chopra:06}, where
the reaction coordinate is a bond-orientational crystalline order
parameter~\cite{steinhardt:83}. A variation of the random-walk strategy to
reconstruct the Helmholtz potential, named umbrella sampling~\cite{torrie:77},
is popular in this context.

A recent alternative is Tethered Monte Carlo~\cite{fernandez:09}, which allows one
to reconstruct the Helmholtz effective potential without random walks in the
space of the effective coordinate. The method is a generalization of Lustig's
microcanonical Monte Carlo~\cite{lustig:98}. One simulates an ensemble where
the reaction coordinate is (almost) constrained to take any desired value. The
associated fluctuation-dissipation formalism permits the accurate reconstruction
of the Helmholtz potential and (if it is so wished) of the canonical
expectation values.  Nevertheless, one is not restricted to a mere speed-up of
a canonical simulation. Very interesting new information can be unearthed from this
different viewpoint.

Furthermore, for a complex enough system, a single reaction coordinate would
not suffice. Wang-Landau, or umbrella sampling simulations are rather
cumbersome if one needs to fine-tune parameters for a two-dimensional random
walk. On the other hand, as we shall show below, a tethered approach to the problem is not
necessarily tougher than the one-dimensional case.

Our aim here is to describe Tethered Monte Carlo, with an emphasis on
its simplicity and generality. We illustrate it with its application
to two challenging problems: hard-spheres crystallization and the
diluted antiferromagnet in an external field (the much easier problem
of the ferromagnetic Ising model was considered in
Refs.~\cite{fernandez:09,martin-mayor:09}). We make emphasis on the
algorithmic aspects of the computation. For a discussion of the
physical results, the reader is referred to
Refs.~\cite{fernandez:11,fernandez:11b,yllanes:11,seoane:12}.  It
turns out that in both problems one needs to compute a Helmholtz
potential depending on two order parameters. However, the
crystallization transition is of the first order, while that of the
diluted antiferromagnet is a continuous one. Hence the strategy
followed in each case differs.

Let us mention as well a tethered computation of the Helmholtz effective
potential for a simple model of glass-forming liquid~\cite{cammarotta:10}.  A
most remarkable feature is that, for this system, an appropriate reaction
coordinate is unknown. However, the problem can be bypassed considering real
replicas.

The layout of the rest of this paper is as follows. Section~\ref{sec:tmc}
describes the general set-up for a tethered simulation using a simple
Metropolis update, in the familiar context of the Ising model. We note {\em en
  passant} that, if a Kasteleyn-Fortuin decomposition is known for the
canonical version of the problem at hand, a cluster update can be implemented
for the Tethered Monte Carlo~\cite{martin-mayor:09}. A detailed derivation of
the tethered formalism is given in Sect.~\ref{sec:tethered-ensemble}.  In
Sect.~\ref{sec:DAFF} we consider the first problem where {\em two} conserved
quantities are needed, the diluted antiferromagnet in a
field. Sect.~\ref{sec:crystallization} describes the application of the
tethered formalism to hard-spheres crystallization. We give our conclusions in
Sect.~\ref{sec:conclusions}.

\section{Tethered Monte Carlo, in a nutshell}\label{sec:tmc}
In this section we give a brief overview of the Tethered Monte Carlo (TMC)
method, including a complete recipe for its implementation in a typical
problem.  This is as simple as performing several independent ordinary MC
simulations for different values of some relevant parameter and then averaging
them with an integral over this parameter.  We shall give the complete
derivations and the detailed construction of the tethered ensemble in
Section~\ref{sec:tethered-ensemble}.

We are interested in the scenario of a system whose phase
space includes several coexisting states, separated by free-energy 
barriers. The first step in a TMC study is identifying the reaction
coordinate $x$ that labels the different relevant phases. This can be 
(but is not limited to) an order parameter. In the remainder of this
section we shall consider a ferromagnetic setting, so the reaction
coordinate will be the magnetization density $m$. 

The goal of a TMC computation is, then, constructing the Helmholtz potential
associated to $m$, $\varOmega_N(\beta,m)$, which will give us all the information about
the system. This involves working in a new statistical ensemble tailored to
the problem at hand, generated from the usual canonical ensemble by Legendre
transformation:
\begin{equation}\label{ec:legendre}
Z_N(\beta,h)= \ee^{N F_N(\beta,h)}=\int \mathrm{d} m \ \mathrm{e}^{N[\beta hm -\varOmega_N(\beta,m)]}\,,
\end{equation}
where $Z_N(\beta,h)$ is
the canonical partition function, $F_N(\beta,h)$ the Gibbs free-energy density
and $N$ is the number of degrees of freedom.

Since in a lattice system the magnetization is discrete, we actually couple it
to a Gaussian bath to generate a smooth parameter, called $\hat m$. The
effects of this bath are integrated out in the formalism.

In order to implement this construction as a workable
Monte Carlo method we need to address two different problems:
\begin{itemize}
\item We need to know how to simulate at fixed $\hat m$. 
\item We need to reconstruct $\varOmega_N(\beta,\hat m)$ from simulations at
  fixed $\hat m$, and afterwards,
  to recover canonical expectation values from
  Eq.~(\ref{ec:legendre}) to any desired accuracy.
\end{itemize}
We shall explain separately how to solve each of the two problems, in the
following two paragraphs.

\subsection{Metropolis simulations in the tethered ensemble}\label{sec:Metropolis}

Let us denote the reaction coordinate by $m$ (for the sake of concreteness let
us think of the magnetization density for an Ising model). The dynamic degrees
of freedom are $\{s_i\}$ (they could be spins, or maybe atomic
positions). Therefore $m$ is a dynamical function (i.e. a function of the
$\{s_i\}$). We wish to simulate at fixed $\hat m$ ($\hat m$ is a {\em
  parameter} closely related to the average value of $m$).

The canonical weight at inverse temperature $\beta$ would be
$\mathrm{exp}[-\beta U]$ where $U$ is the interaction energy. Instead, the
tethered weight is (see Ref.~\cite{fernandez:09} and
Sect.~\ref{sec:tethered-ensemble} for a derivation)
\begin{equation}\label{ec:weight1}
\omega_N(\beta,\hat m;\{s_i\})=\mathrm{e}^{-\beta U + N (m-\hat m)} (\hat
m -m)^{(N-2)/2} \theta(\hat m -m)\,.
\end{equation}
The Heaviside step function $\theta(\hat m -m)$ imposes the constraint that
$\hat m> m(\{s_i\})$.

The tethered simulations with weight (\ref{ec:weight1}) are exactly like a
standard canonical Monte Carlo in every way (and satisfy detailed balance,
etc.). For instance, in an Ising model setting, the common Metropolis
algorithm~\cite{metropolis:53} is
\begin{enumerate}
\item Select a spin $s_i$.
\item The proposed change is flipping the spin, $s_i\to -s_i$.~\footnote{In an
  atomistic simulation, one would try to displace a particle, or maybe to
  change the volume of the simulation box.}
\item The change is accepted with probability\footnote{In general,
the Metropolis probability has to take into account both 
the weight and the probability of proposing this particular
change~\cite{landau:05}. However, for this simple
problem the latter is symmetric, so it quotients out.}
\begin{equation}
P(s_i \to -s_i) = \min \{ 1,\omega_\mathrm{new}/\omega_\mathrm{old}\}.
\end{equation}
\item Select a new spin $s_i$ and repeat the process.
\end{enumerate}

We remark that the above outlined algorithm produces a Markov chain entirely
analogous to that of a standard, canonical Metropolis simulation. As such it has
all the requisite properties of  a Monte Carlo simulation (mainly reversibility
and ergodicity). Tethered
mean values can be computed as the time average along the simulation of the
corresponding dynamical functions (such as internal energy, magnetization
density, etc.). Statistical errors and autocorrelation times can be computed
with standard techniques~\cite{amit:05,sokal:97}.

The actual magnetization density is constrained (tethered) in this simulation,
but it has some leeway (the Gaussian bath can absorb small variations in $m$,
see Sect.~\ref{sec:tethered-ensemble}). In fact, its fluctuations are
crucial to compute an important dynamic function, whose introduction would
seem completely unmotivated from a canonical-ensemble point of view:
\begin{equation}\label{eq:campo-tethered}
\hat b \equiv -\frac{1}{N} \frac{\partial \log \omega_N(\beta,\hat
  m;\{s_i\})}{\partial \hat m}= 1 - \frac{N-2}{2 N [\hat m - m(\{s_i\})]}\,.
\end{equation}
One of the main goals of a tethered simulation is the accurate computation of
the expectation value $\langle \hat b\rangle_{\hat m}$.

The case where one wishes to consider two reaction coordinates $m_1$ and $m_2$
is completely analogous:
\begin{eqnarray}
\omega_N(\beta,\hat m_1,\hat m_2;\{s_i\})&=&\mathrm{e}^{-\beta U + N (m_1-\hat
  m_1)+ N (m_2-\hat m_2) } \times \\\nonumber &\times&(\hat m_1 -m_1)^{(N-2)/2} (\hat m_2 -m_2)^{(N-2)/2}
\theta(\hat m_1 -m_1) \theta(\hat m_2 -m_2)\,,\\
\hat b_1&\equiv& -\frac{1}{N} \frac{\partial \log \omega_N(\beta,\hat
  m_,\hat m_2;\{s_i\})}{\partial \hat m_1}= 1 - \frac{N-2}{2 N [\hat m_1 -
    m_1(\{s_i\})]}\,,\\
\hat b_2&\equiv& -\frac{1}{N} \frac{\partial \log \omega_N(\beta,\hat
  m_,\hat m_2;\{s_i\})}{\partial \hat m_2}= 1 - \frac{N-2}{2 N [\hat m_2 - m_2(\{s_i\})]}\,.
\end{eqnarray} 

We note as well that, in the context of supercooled liquids 
crystallization~\cite{tenwolde:95}, a tethered weight slightly different from
Eq.~(\ref{ec:weight1}) has been in use (however, the importance  of the dynamic
function $\hat b$ was apparently not recognized):
\begin{eqnarray}\label{ec:weight2}
\omega_N(\beta,\hat m;\{s_i\})&=&\mathrm{e}^{-\beta U - N \alpha (\hat
    m- m)^2/2}\,,\\
\hat b &=& \alpha (\hat m - m)\,.
\end{eqnarray}
Here, $\alpha$ is a tunable parameter that can be handy to control the
deviations of $\hat m$ from $m$~\cite{fernandez:11}. Note as well
that the weight~\eqref{ec:weight2} does {\em not} impose $\hat m > m$.

For the Ising model, a Metropolis Tethered Monte Carlo simulation 
reconstructs the crucial tethered magnetic field $\hat b$ 
without critical slowing down
(see \cite{fernandez:09} for a
benchmarking study). This may be considered surprising for what is 
a local update algorithm, but notice that the constraint on $\hat m$
is imposed globally. Non-magnetic observables, such as the energy,
do not enjoy this non-local information and hence show a typical $z\approx2$
critical slowing down (although the correlation times are low enough to
permit equilibration for very large systems,~\cite{fernandez:09}).

Let us stress that the above outlined update algorithm is by no means
the only one possible. For instance, the Fortuin-Kasteleyn construction~\cite{kasteleyn:69,fortuin:72}
can be performed just as easily in the tethered ensemble, so 
we can consider tethered simulations with cluster
update methods~\cite{swendsen:87,wolff:89,edwards:88}. This is demonstrated
in~\cite{martin-mayor:09}, where the tethered version of the Swendsen-Wang
algorithm is shown to have the same critical slowing down
as the canonical one for the $D=3$ Ising model ($z\approx0.47$).
This is an example that the use of the tethered formalism implies
no constraints on the choice of Monte Carlo algorithm, nor hinders it in the 
case of an optimized method.

\subsection{Reconstructing the Helmholtz effective potential from simulations
  at fixed $\hat m$}

\begin{figure}
\centering
\includegraphics[height=0.7\linewidth,angle=270]{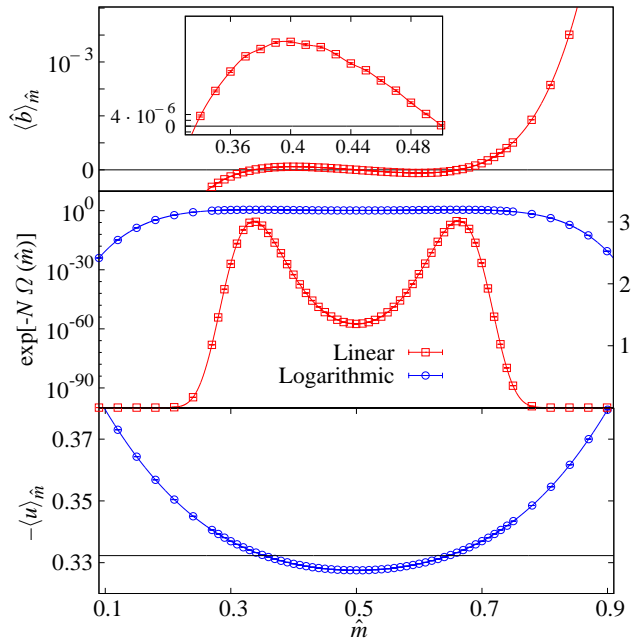}
\caption{(Color online) Computation of the Helmholtz potential $\Omega_N$
and the canonical expectation values from
tethered averages in a $D=3$, $L=64$ ferromagnetic Ising model
at the critical temperature. The upper panel shows the tethered magnetic field
$\langle \hat b\rangle_{\hat m}$, with an inset zooming in on
the region between its two zeros. The statistical errors cannot be seen at this scale
(except for the leftmost points in the inset). The integral of this quantity 
is the Helmholtz potential $\Omega_N(\hat m)$.
The middle panel shows the $p(\hat m)=\exp[-N\Omega_N(\hat m)]$
in a linear (right axis) and in a logarithmic scale (left axis).
Finally, the bottom panel shows the tethered expectation values
of the energy density $u$. Their integral over the whole 
$\hat m$  range, weighted with $p(\hat m)$, gives
the canonical expectation value $\langle u \rangle = -0.3322894(36)$
(horizontal line). See Ref.~\cite{martin-mayor:09} for further details on
these simulations.
}
\label{fig:tethered}
\end{figure}

The steps in a TMC simulation are, then, (see also Figure~\ref{fig:tethered})
\begin{enumerate}
\item Identify the range of $\hat m$ that covers 
the relevant region of phase space. 
Select $N_{\hat m}$ points $\hat m_i$, evenly spaced 
along this region.
\item For each $\hat m_i$ perform a Monte Carlo simulation where the smooth reaction
  parameter $\hat m$ will be fixed at $\hat m=\hat m_i$.
\item We now have all the relevant physical observables
as discretized functions of $\hat m$. We 
denote these tethered averages at fixed $\hat m$ by
$\langle O\rangle_{\hat m}$. 
\item The average values in the canonical ensemble, denoted by
 $\langle O\rangle$, can be recovered with a simple integration
\begin{equation}
\langle O \rangle = \int_{\hat m_\text{min}}^{\hat m_\text{max}} \dd \hat m\
p(\hat m) \langle O\rangle_{\hat m}.
\end{equation}
In this equation the probability density $p(\hat m)$ is
\begin{equation}
p(\hat m) = \ee^{-N \Omega_N(\hat m)},\quad \Omega_N(\hat m) = \Omega_N(\hat m_\mathrm{min})+ 
\int_{\hat m_\text{min}}^{\hat m} \dd \hat m'\
\langle \hat b\rangle_{\hat m'}.
\end{equation}
The tethered field $\langle \hat b\rangle_{\hat m}$ was defined in
Eq.~\eqref{eq:campo-tethered}.  The integration constant $\Omega_N(\hat
m_\mathrm{min})$ is chosen so that the probability is normalized.
\item If we are interested in canonical averages in the presence
of an external magnetic field $h$, we do not have to recompute
the $\langle O\rangle_{\hat m}$, only $\Omega_N$.
This is as simple as shifting the tethered magnetic field:
$\langle \hat b\rangle_{\hat m} \to \langle \hat b\rangle_{\hat m} - \beta h$.
\item In order to improve the precision and avoid systematic errors, 
we can run additional simulations in the region where $p(\hat m)$
is largest.
\end{enumerate}
The whole process is illustrated in Figure~\ref{fig:tethered}, where we compute
the energy density at the critical temperature in an $L=64$ lattice of 
the $D=3$ Ising model. Notice that the tethered expectation values $\langle u\rangle_{\hat m}$
vary in about 
$10\%$ in our $\hat m$ range, but the computation of the effective potential
is so precise that the averaged value for the energy, $\langle u \rangle = -0.3322894(36)$, 
has a relative error of only $\sim10^{-5}$.

This is the general TMC algorithm for the computation of canonical averages
from the Helmholtz potential. As we shall see in some of the applications,
sometimes the integration over all phase space in step 4 is not needed and one
can use the ensemble equivalence property to recover the $\langle O \rangle$
from the $\langle O\rangle_{\hat m}$ through saddle-point equations, remember
Eq.~\eqref{ec:legendre}.  In other words, the tethered averages can be
physically meaningful by themselves. We note as well that our crystallization
study in Sect.~\ref{sec:crystallization} is built entirely over the
effective-potential, one never uses the $p(\hat m)$.

As will be shown in the next section, the reconstruction of canonical
averages from the combination of tethered averages does not
involve any approximation. We can achieve any desired accuracy, provided
we use a sufficiently dense grid in $\hat m$ (to control systematic errors)
and simulate each point for a sufficiently long time (to reduce
statistical ones). Table~\ref{tab:energies} shows the kind of 
precisions that we can achieve. One could initially think that 
the computation of the exponential in $p(\hat m) = \exp[-N \Omega_N(\hat m)]$
would produce unstable or imprecise results for large system sizes.
Instead, the combination  of 
self-averaging and no critical slowing down makes the numerical precision
grow with $N$.

\begin{table}
\centering
\caption{Energy density of the $D=3$ ferromagnetic Ising
model computed with the Tethered Monte Carlo method, showing 
that the reconstruction of canonical averages can be performed
with great accuracy. The second column shows the  number
of points in the $\hat m$ grid and the third the number
of Monte Carlo sweeps taken on each (we use a cluster
update scheme). See~\cite{martin-mayor:09}
for further details on these simulations.}
\begin{tabular*}{\columnwidth}{@{\extracolsep{\fill}}rrcl}
\hline
$L$ & $N_{\hat m}$ & MCS &\multicolumn{1}{c}{$-\langle u\rangle$}\\
\hline
16 & 91  & $10^6$ & 0.344905(35) \\ 
32 & 91  & $10^6$ & 0.335730(26) \\
64 & 109 & $10^6$ & 0.3322894(36) \\
128 & 50 & $10^6$ & 0.3309831(15) \\
\hline
\end{tabular*}
\label{tab:energies}
\end{table}

\section{The tethered ensemble}\label{sec:tethered-ensemble}
In this section we construct the statistical ensemble that supports 
the TMC method. We shall consider the case of the $D$-dimensional
ferromagnetic Ising model. Reproducing the steps of this 
derivation for any other system is straightforward.  We include
details on how to introduce several tethered variables (subsection~\ref{sec:several-tethers}).
Subsection~\ref{sec:ensemble-equivalence} discusses the relationship between
tethered and canonical expectation values from the point
of view of the ensemble equivalence property.

The Ising model is characterized by the following partition function
\begin{equation}
Z = \sum_{\{s_\bx\}} \exp\biggl[ \beta  \sum_{\langle\bx,\by\rangle} 
s_\bx s_\by\biggr],
\end{equation}
where the angle brackets indicate that the sum is restricted to first
neighbors and the spins are $s_\bx=\pm1$. In this section we will be working
at fixed $\beta$. Hence, to lighten the expressions we shall drop the explicit
$\beta$ dependencies.

The energy and magnetization of this system are
\begin{align}
U &= Nu = - \sum_{\langle\bx,\by\rangle} s_\bx s_\by,&
M &= Nm = \sum_\bx s_\bx,
\end{align}
($N=L^D$ is the number of spins in the system).

The canonical average of a generic observable $O$ is
\begin{align}
\langle O \rangle = \frac{1}{Z} \sum_{\{s_\bx\}} O(\{s_i\})\  \ee^{-\beta U}.
\end{align}
Since this is a ferromagnetic system, we may be interested
in considering the average value of $O$ conditioned to different 
magnetization regions. The naive way of doing this  would 
be 
\begin{align}
\langle O | m \rangle = \frac{\bigl\langle O \delta\bigl(m- \sum_\bx s_\bx/N\bigr)\bigr\rangle}
{\bigl\langle\delta\bigl(m- \sum_\bx s_\bx/N\bigr)\bigr\rangle}
\end{align}
The canonical average could then be recovered by a weighted
average of the $\langle O|m\rangle$,
\begin{equation}\label{eq:O-m-fijo}
\langle O \rangle = \sum_m \langle O|m\rangle p_1(m),\qquad
p_1(m)= {\biggl\langle\delta\biggl(m- \sum_\bx s_\bx/N\biggr)\biggr\rangle}
\end{equation}
In the thermodynamical limit, the reaction coordinate $m$ varies
continuously from $-1$ to $1$.  For a finite system, however, there
are only $N+1$ possible values of $m$, so $p_1(m)$ is a comb-like
function.

We want to construct a statistical ensemble where a smooth Helmholtz potential
can be defined in finite lattices. 
In order to do this, the first step is extending the configuration space 
with a bath $R$ of $N$ Gaussian demons and defining the smooth magnetization $\hat M$,
\begin{equation}
\hat M = N \hat m= M + R.
\end{equation}
These demons can be introduced either linearly or quadratically,\footnote{In fact, 
they need not even be Gaussian variables, although we shall not consider
other distributions here.}
\begin{align}
R^{(\text{Q})} &= \sum_i \eta_i^2/2,\label{eq:demonios-cuadraticos}	\\ 
R^{(\text{L})} &= \sum_i \eta_i.\label{eq:demonios-lineales}
\end{align}
Now our partition function is
\begin{align}
Z &= \int_{-\infty}^\infty \biggl(\prod_{i=1}^N \frac{\dd\eta_i}{\sqrt{2\pi}}\biggr)
\sum_{\{s_\bx\}} \exp\biggl[ -\beta U - \sum_i \eta^2_i/2 \biggr],\\
p_2(r) &= \int_{-\infty}^\infty \biggl(\prod_{i=1}^N \frac{\dd\eta_i}{\sqrt{2\pi}}\biggr)
\exp\biggl[- \sum_i \eta^2_i/2\biggr]
\delta(r - R/N)
\end{align}
Notice that the demons are statistically independent from the spins.
The convolution of $p_1(m)$ and $p_2(r)$ then gives the probability 
density function for $\hat m$,
\begin{equation}\label{eq:convolucion}
p(\hat m) = \int\dd m \dd r\ p_1(m) p_2(r)
\delta(\hat m-m-r).
\end{equation}
So $p(\hat m)$ is essentially a smooth version of 
$p_1(m)$. We can do an analogous construction for off-lattice systems. 
In this case, the demons would increase the dispersion
in the already continuous $p_1(x)$. We can control these fluctuations
by considering a tunable number of demons, see~\eqref{eq:sum-demons-lineales} and
Section~\ref{sec:crystallization}.

We can now rewrite the canonical average $\langle O\rangle$ as
\begin{equation}\label{eq:canonical-average}
\langle O \rangle = \int\dd \hat m\  \langle O\rangle_{\hat m} 
\ p(\hat m),
\end{equation}
where the tethered average $\langle O\rangle_{\hat m}$ is
\begin{equation}
\langle O \rangle_{\hat m}= \frac{1}{Z p(\hat m)} \int_{-\infty}^\infty
\prod_{i=1}^N \dd \eta_i \sum_{s_\bx} O(\{s_\bx\})\
\ee^{-\beta U - \sum_i \eta^2_i/2} \delta(\hat m - m - R/N).
\end{equation}
This way, each $\langle O\rangle_{\hat m}$ 
will have contributions from
all the $\langle O|m\rangle$, with a weight that will depend on the distance between
$m$ and $\hat m$. This generates smooth $\langle O\rangle_{\hat m}$
and $p(\hat m)$, so we can now define the Helmholtz effective potential
in our finite lattice as
\begin{equation}
p(\hat m) = \ee^{-N\Omega_N(\hat m)}
\end{equation}
It may seem that with our extended configuration space, 
we have sacrificed too much in order to get
smooth functions, but we can actually integrate the demons out 
with the Dirac deltas:
\begin{equation}
\ee^{-N\Omega_N(\hat m)} = \frac1Z\sum_{\{s_\bx\}} \omega_N(\hat m,\{s_\bx\}),
\end{equation}
where the weight $\omega_N(\hat m ,\{s_\bx\})$ is, for our linear and quadratic
demons (neglecting irrelevant constant factors),
\begin{align}\label{eq:omega}
\omega^{(\text{Q})}_N (\hat m,\{s_\bx\}) &\propto \ee^{-\beta U + M-\hat M} (\hat m-m)^{(N-2)/2} \theta(\hat m-m)\\
\omega^{(\text{L})}_N (\hat m, \{s_\bx\}) &\propto \ee^{-\beta U -(M-\hat M)^2/(2N)}.
\end{align}
In general, $\omega_N(\hat m,\{s_\bx\})$ will be of the following form,
\begin{equation}\label{eq:omega-gamma}
\omega_N(\hat m,\{s_\bx\}) \propto \ee^{-\beta U} \gamma(\hat m,m),
\end{equation}
that is, the thermal component typical of a canonical simulation and
a magnetic factor that depends only on $m$ and $\hat m$.
The weight $\omega_N$ defines a new statistical ensemble, 
so we can now rewrite the tethered averages as
\begin{equation}
\langle O\rangle_{\hat m} = \frac{ \sum_{\{s_\bx\}} O(\{s_\bx\}) 
\omega_N(\hat m,\{s_\bx\})}{ \sum_{\{s_\bx\}} 
\omega_N(\hat m,\{s_\bx\})}
\end{equation}
A particularly important tethered average is the $\hat m$-derivative
of the effective potential, the tethered magnetic field $\langle \hat b\rangle_{\hat m}$,
\begin{equation}\label{eq:hatb}
\frac{\partial \Omega_N}{\partial \hat m} = \langle \hat b\rangle_{\hat m}.
\end{equation}

For our two kinds of demons this is
\begin{align}
\hat b^{(\text{Q})} &= 1-\frac{1/2-1/N}{\hat m-m} ,\\
\hat b^{(\text{L})} &= \hat  m - m.
\end{align} 

Definition~\eqref{eq:hatb}, together with the condition that $p(\hat m)= \ee^{- N\Omega_N(\hat m)}$ be normalized,
allows us to reconstruct the Helmholtz potential from tethered averages
alone. The canonical averages can then be recovered through 
Eq.~\eqref{eq:canonical-average}. 

The tethered magnetic field is essentially a measure of the fluctuations
in $m$, which illustrates the dual roles the magnetization and magnetic
field play in the canonical and tethered formalism.  This
is  further demonstrated by the tethered fluctuation-dissipation formula
\begin{equation}\label{eq:fluctuation-dissipation}
\frac{\partial \langle O\rangle_{\hat m}}{\partial \hat m}
= \left\langle \frac{\partial O}{\partial \hat m}\right\rangle_{\hat m}
+ N [ \langle O \hat b\rangle_{\hat m} - \langle O \rangle_{\hat m}
\langle \hat b\rangle_{\hat m}].
\end{equation}

Notice that the values of $\hat m$ where $\langle \hat b\rangle_{\hat m}=0$
will define maxima and minima of the effective potential and, hence,
of the probability $p(\hat m)$.

The tethered ensemble is, thus, constructed via a Legendre transformation 
of the canonical one, that exchanges the roles of the magnetic field
and the magnetization. In fact, we can write the canonical partition 
for a given value of the applied magnetic field in the 
following way
\begin{equation}\label{eq:partition-function-h}
Z(h) = \ee^{N F_N(h)} = \int \dd\hat m\ \ee^{-N [\Omega_N(h) - \beta h \hat m]}.
\end{equation}
One interesting consequence is that computing canonical
averages for non-zero magnetic fields does not imply 
changing (recomputing) the tethered averages, only their weights,
\begin{equation}\label{eq:canonical-average-h}
\langle O \rangle (h) = \frac{1}{Z(h)} \int \dd \hat m \ \langle O\rangle_{\hat m}
\ \ee^{-N [\Omega_N(h) - \beta h \hat m]},
\end{equation}
in other words, we just have to shift the tethered magnetic field,
$\langle \hat b\rangle_{\hat m} \to \langle \hat b \rangle_{\hat m} - \beta h$.

A final general point concerns the introduction of the Gaussian demons. 
Our definitions~\eqref{eq:demonios-cuadraticos} and \eqref{eq:demonios-lineales}
and subsequent construction make the assumption that there 
are as many demons as spins. While this choice seems natural, it is by no means
a necessity. In fact, in order to control the fluctuations in the reaction
coordinate, it may in occasion be interesting to consider a variable 
number of demons (see Section~\ref{sec:crystallization}):
\begin{align}\label{eq:sum-demons-lineales}
R_\alpha & =\frac1\alpha \sum_{i=1}^{\alpha N} \eta_i,&
\hat M = N \hat m = M + R_\alpha,
\end{align}
where $\alpha$ is the tunable parameter. All the computations in this section
can be easily reconstructed for this more general case. As an example, 
the tethered magnetic field with $\alpha N$ linearly added demons is
\begin{equation}\label{eq:b_lineal}
\hat b^{(\text{L})} = \alpha(\hat m- m).
\end{equation}

\subsection{Several tethered variables}\label{sec:several-tethers}
Throughout this section we have considered an ensemble with only one tethered
quantity. However, as we shall see in the following sections, it is often 
appropriate to consider several reaction coordinates at the same time. 
The construction of the tethered ensemble in such a study presents no 
difficulties. We start by coupling reaction coordinates $x_i$, $i=1..n$,  
with $N$ demons each,
\begin{align}
\hat X_1 =N\hat x_1=  X_1 + R_1,\quad \ldots,\quad
\hat X_n =N\hat x_n=  X_n + R_n.
\end{align}
We then follow the same steps of the previous section, with the consequence
that the tethered magnetic field is now a conservative field computed
from an $n$-dimensional potential $\Omega_N(\hat{\bx})$
\begin{align}
\boldsymbol\nabla \Omega_N &= \bigl( \partial_{\hat x_1} \Omega_N,\ldots,
\partial_{ \hat x_n} \Omega_N\bigr) = \hat{\boldsymbol B}\\
\hat {\boldsymbol B } &= (\langle \hat b_1\rangle_{\hat{\boldsymbol x}},\ldots,
\langle \hat b_n\rangle_{\hat{\boldsymbol x}}),
\end{align}
and each of the $\hat b_i$ is of the same form as in the case with only 
one tethered variable. Similarly, the tethered weight of Eq.~\eqref{eq:omega-gamma} is now
\begin{equation}
\omega^{(n)} (\hat{\boldsymbol x}, \{s_\bx\}) = \ee^{-\beta U}
\gamma(\hat x_1,x_1) \gamma(\hat x_2,x_2)\cdots\gamma(\hat x_n,x_n).
\end{equation}

\subsection{Ensemble equivalence and spontaneous symmetry breaking}\label{sec:ensemble-equivalence}
Throughout this section we have implicitly assumed that the final goal
of a Tethered Monte Carlo computation is eventually to reconstruct the canonical
averages. As we shall 
see in some of the applications, however, this is not always the case.
One example is the computation of the hyperscaling violations exponent
of the RFIM in~\cite{fernandez:11b}, which is made directly from $\Omega_N$.

Still, most of the time the averages in the canonical ensemble are the
ones with a more direct physical interpretation (fixed temperature,
fixed applied field, etc.). In principle, their computation implies
reconstructing the whole effective potential $\Omega_N$ and using it
to integrate over the whole coordinate space, as
in~\eqref{eq:canonical-average} or, more generally,
\eqref{eq:canonical-average-h}. Sometimes, however, the connection
between the tethered and canonical ensembles is easier to make.  Let
us return to our ferromagnetic example, with a single tethered
quantity $m$.  Recalling the expression of the canonical partition
function in terms of $\Omega_N$ for finite $h$,
Eq.~\eqref{eq:partition-function-h}, a steepest-descent tells us
that, for large $N$, the integral is dominated by one saddle point
such that
\begin{equation}
\partial_{\hat m} [\Omega_N(\hat m) - \beta h \hat m] = 0\qquad\Longrightarrow
\qquad \langle \hat b\rangle_{\hat m} = \beta h.
\end{equation}
Clearly, this saddle point (actually, the global minimum of the
effective potential) rapidly grows in importance with the system size
$N$, to the point that we can write
\begin{equation}\label{eq:ensemble-equivalence}
\lim_{N\to \infty} \langle O \rangle(h) =\lim_{N\to \infty} \langle O \rangle_{\hat m(h)}.
\end{equation}
That is, in the thermodynamical limit we can identify the canonical average
for a given applied magnetic field $h$ with the tethered average computed 
at the saddle point defined by $h$ (which is nothing more than the point
where the tethered magnetic field coincides with the applied magnetic
field). 

This ensemble equivalence property would be little more than a curiosity if
it were not for the fact that the convergence is actually extremely fast
(see \cite{fernandez:09} for a study). Therefore, in many practical
applications the equivalence~\eqref{eq:ensemble-equivalence} can 
be made even for finite lattices (see Section~\ref{sec:DAFF} below for 
an example).

The saddle-point approach can be applied to systems with spontaneous symmetry
breaking. In the typical analysis, one has to perform first a large-$N$ limit and then a small-field
one (this is troublesome for numerical work, where one usually has to 
consider non-analytical observables such as $|m|$). In the tethered ensemble 
we can implement this double limit in an elegant way by considering a restricted
range in the reaction coordinate from the outset (see~\cite{fernandez:09} for a 
straightforward example in the Ising model and Section~\ref{sec:DAFF-disorder-average}
for a more subtle one).

\section{The diluted antiferromagnet in a magnetic field}\label{sec:DAFF}
In this section we consider the application of the tethered formalism 
to the diluted antiferromagnet in an external field (DAFF), one of the simplest 
physical systems showing cooperative behavior in the presence of quenched
disorder. A good experimental introduction to this system can be
found in~\cite{belanger:97}. It was shown in~\cite{fishman:79,cardy:84} that
the DAFF belongs to the same universality 
class as the random field Ising model (RFIM),
a system more amenable to analytical
treatment (see \cite{nattermann:97} for a review and~\cite{dedominicis:06}
for a field-theoretical approach). 

Experimental, Monte Carlo and analytical studies
of the DAFF/RFIM agree that this system experiences a phase transition for 
more than two spatial dimensions.  At low temperatures, if the applied 
magnetic field  $h$ is small, the DAFF remains in an antiferromagnetically
ordered state. If the magnetic field (or the temperature) is raised,
however, the ferromagnetic 
part of the Hamiltonian weakens the antiferromagnetic correlations and 
the DAFF crosses into a paramagnetic state. This transition occurs
across a smooth critical surface in the $(\beta,h,p)$ space ($p$
characterizes the dilution of the system).

That much is generally accepted. However, despite  decades of interest 
on this system, most  of the details of this phase transition remain 
controversial. For instance, the analytical work on the RFIM 
favors a second-order scenario. Most  recent numerical and experimental
studies of the DAFF agree~\cite{vink:10,hartmann:01,slanic:99,ye:02}, but 
some recent work~\cite{sourlas:99,maiorano:07} has found signatures
characteristic of first-order transitions (such as metastability, 
or a very low critical exponent $\beta$ that raises the possibility
of a discontinuous transition). These apparent inconsistencies
between numerical work and analytical predictions have even
led some researchers  to consider the possibility that
the RFIM and DAFF may not be equivalent after all. 
Even allowing for this possibility,
the comparison of numerical and experimental
work on the DAFF has not always yielded completely 
satisfactory results. For instance, some experiments have reported
a divergent specific heat~\cite{belanger:83,belanger:98}, not observed numerically~\cite{hartmann:01}.

The problems on the numerical front can be understood if we 
notice that the DAFF is particularly ill-suited to 
numerical investigation in the canonical ensemble. On the one
hand, the system suffers from a very severe thermally activated 
critical slowing down, where the characteristic times 
grow as $\log \tau \sim \xi^{\theta}$, with $\theta\approx 1.5$.
On the other hand, the statistical averages in the canonical ensemble
are dominated by extremely rare events, which gives rise
to strong violations of self-averaging~\cite{parisi:02}.

Finally, some crucial physical results are hardly accessible 
to a canonical computation. Perhaps the most important 
is the hyperscaling violations exponent $\theta$. This is an 
additional critical exponent that one must introduce to 
make the theory consistent, and that modifies the 
standard hyperscaling relation, $2-\alpha=D \nu$, in the following
way,
\begin{equation}\label{eq:hyperscaling}
2-\alpha = \nu (D-\theta).
\end{equation}
In the canonical ensemble, 
the exponent $\theta$ is associated to the free-energy barrier
between the ordered and the disordered phase~\cite{vink:10,fischer:11}, $\Delta F\propto L^\theta$.
Hence, it is
a good marker to distinguish a first-order scenario ($\theta\geq D-1$)
from a second-order one ($\theta<D-1$). At the same time, it is
extremely difficult to compute with standard Monte Carlo methods.

In~\cite{fernandez:11b} we performed a Tethered Monte Carlo
study of this system and found that many of the problems with 
previous numerical work were solved. As one wishes to ascertain that a
first-order transition with the magnetic field is {\em not} present, it is
only natural to tether the conjugate magnitude, the standard magnetization. On
the other hand, the largest barriers are associated to the order parameter
(the staggered magnetization, see below), which was also tethered.

A consistent picture of the phase transition, which was shown to be
continuous, emerged.  Self-averaging was restored and thermalization greatly
accelerated.  We were able to obtain the three independent critical exponents
and found a (slow) divergence of the specific heat, in accordance with
experiments. In particular, the critical exponent~$\theta$, so elusive to
canonical computations, was determined with some precision: $\theta = 1.47(2)$
(the free-energy barriers related to $\theta$ are immediately translated into
effective-potential barriers, readily computed with the methods of the
previous sections).

In the remainder of this section we shall explain in some detail
the application of the tethered formalism to this problem.
We first define the model and some basic observables in~\ref{sec:DAFF-model}.
This is followed in~\ref{sec:DAFF-canonical}
by an examination of the DAFF with traditional (canonical) 
methods, that expands on the problems discussed above.
Next, we construct a tethered formalism adapted to the DAFF
and discuss how to restore self-averaging to the disorder
averages. In~\ref{sec:DAFF-simulations} we report the parameters
of our simulations and in~\ref{sec:DAFF-thermalization}
we discuss their thermalization and possible optimizations.
Finally, in~\ref{sec:DAFF-geometry} we include a geometrical
investigation of the instanton configurations,
not included in~\cite{fernandez:11b}.
This serves as a good example of how a tethered computation is not restricted
to reproducing canonical results more efficiently, but can access interesting
information that is completely hidden from a canonical study.

\subsection{Model and observables}\label{sec:DAFF-model}
We consider the diluted antiferromagnet in a field (DAFF) on a square lattice
with $N=L^3$ nodes and periodic boundary conditions
\begin{equation}\label{eq:DAFF}
\mathcal H = \sum_{\langle \boldsymbol x , \boldsymbol y\rangle} \epsilon_\bx s_\bx
\epsilon_\by s_\by - h \sum_\bx \epsilon_\bx s_\bx.
\end{equation}
In addition to the Ising spins ($s_\bx=\pm 1$), the Hamiltonian includes
occupation variables $\epsilon_\bx$. These are randomly 
distributed according to $P(\epsilon) = p \delta(\epsilon-1) + (1-p) \delta(\epsilon)$.
Throughout this paper we shall consider a spin concentration of $p=0.7$. This 
value is very far from the percolation threshold of $p\approx0.31$~\cite{stauffer:84},
but not so high as to be dominated by the pure case ($p=1$). The $\epsilon_\bx$ 
are quenched variables, which means that we should perform first the thermal
average for each choice of the $\{\epsilon_\bx\}$ and only afterwards compute
the disorder average.  In the canonical ensemble, we shall denote
the thermal average  by $\langle\cdots\rangle$
and the disorder average by $\overline{(\cdots)}$.

For each spin configuration, we can define the standard and staggered magnetizations
\begin{eqnarray}
M &=& Nm = \sum_\bx \epsilon_\bx s_\bx, \label{eq:M}\\
\Ms &=& N m_\mathrm{s} = \sum_\bx \epsilon_\bx s_\bx  \pi_\bx,
\quad \pi_\bx=\ee^{\ii \pi \sum_{\mu=1}^D x_\mu} \label{eq:Ms}.
\end{eqnarray}
We shall distinguish between the energy $E$ and the first-neighbors connectivity $U$,
\begin{eqnarray}
E &=& N e = \sum_{\langle \boldsymbol x , \boldsymbol y\rangle} \epsilon_\bx s_\bx
\epsilon_\by s_\by - h \sum_\bx \epsilon_\bx s_\bx,\label{eq:E}\\
U &=& N u = \sum_{\langle \boldsymbol x , \boldsymbol y\rangle} \epsilon_\bx s_\bx
\epsilon_\by s_\by. \label{eq:U}
\end{eqnarray}
\subsection{The DAFF in canonical Monte Carlo simulations}\label{sec:DAFF-canonical}
In this section we shall explain the limitations inherent in a canonical
study and demonstrate how the tethered formalism can be used 
to gain more information about the system. In order to do so, we 
shall first carry out a comparative study of a single disorder realization
(sample) in both the canonical and tethered ensembles. We shall 
then explain how to perform the disorder average in the tethered 
formalism and provide the motivation for the approach taken in this study.

\begin{figure}
\centering
\includegraphics[height=.9\columnwidth,angle=270]{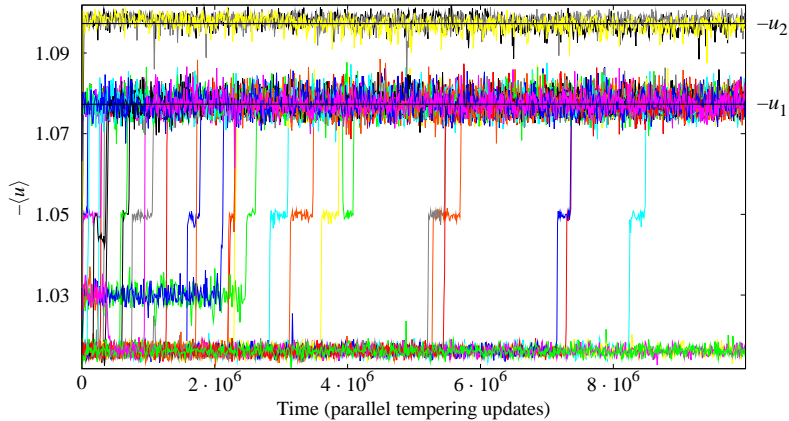}
\caption{(Color online) Monte Carlo histories of the first-neighbor
  connectivity density $u$ for different runs of the same sample in a
  canonical parallel tempering simulation ($L=24$, $T=1.6, h=2.4,
  p=0.7$).  After a variable simulation time, the runs eventually
  reach one of two local minima (labeled by $u_1$ and $u_2$), but the
  simulations are not long enough for us to see any tunneling between
  the two.}\label{fig:historia_canonico}
\end{figure}
We have performed several
parallel-tempering~\cite{hukushima:96,marinari:98b} simulations of the
same disorder realization for an $L=24$ system.\footnote{We used $40$
  evenly spaced temperatures in the range $1.6\leq T\leq 2.575$,
  choosing $h$ so that $h/T=1.5$.  The dilution was $p=0.7$. These
  parameters are taken from~\cite{maiorano:07}, although in that
  reference even lower temperatures are explored.}  In
Figure~\ref{fig:historia_canonico} we can see how after some time the
systems reach one of two local minima of the effective potential, with
first-neighbor connectivities, Eq.~(\ref{eq:U}) $u^{(1)}=-1.07735(8)$
and $u^{(2)}=-1.0973(7)$. The corresponding energy densities,
Eq.~(\ref{eq:E}), are $e^{(1)}=-1.37608(2)$ and $e^{(2)}=-1.3820(2)$.

On the basis of these data, one would think that 
these two minima are relevant to the equilibrium of this particular sample, so that
if we wait a sufficiently long time we will  eventually see the system tunnel
back and forth between the two of them.
This apparent metastability could lead to the interpretation that the system is
experiencing a first-order phase transition. There are two problems, however.
The first one is that we have not actually seen
the tunneling in any of our simulations.\footnote{We have run $100$ thermal histories taking
$10^7$ parallel-tempering steps in each, which suggests that the 
tunneling probability is upper bounded by $10^{-9}$.} In other words, canonical parallel-tempering is
not able to thermalize the system in a reasonable amount of time. Notice that obtaining the two
minima in separate simulations is not enough, we need to see 
the tunneling in order to know their relative weight.

\begin{figure}
\centering
\includegraphics[height=.7\columnwidth,angle=270]{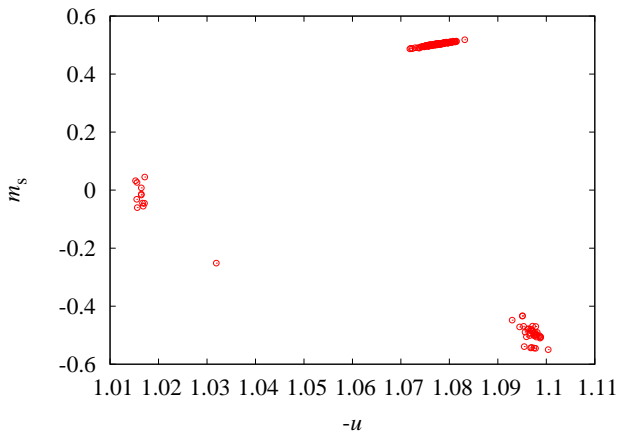}
\caption{(Color online) Scatter plot of the first-neighbor
  connectivity and the staggered magnetization, as computed on the
  very last $0.1\%$ of the Monte Carlo history for the simulations in
  Figure~\ref{fig:historia_canonico}. Most of the simulations have
  reached one of two local minima with opposite staggered
  magnetization.}\label{fig:scatter_canonico}
\end{figure}

The second problem is illustrated on Figure~\ref{fig:scatter_canonico}.  In it
we represent a scatter plot of the staggered magnetization against $u$ for the
runs in Figure~\ref{fig:historia_canonico}. It is readily apparent that the
two local minima correspond to antiferromagnetic systems with opposite sign
of the order parameter ($m_\text{s}^{(1)}=0.5023(3)$,
$m_\text{s}^{(2)}=-0.543(3))$, so the observed metastability does not
correspond to the disordered-antiferromagnetic transition in which we are
interested.\footnote{Notice that an individual sample of the DAFF does not
  have the usual $Z_2$ symmetry in antiferromagnetic systems, since the number
  of spins aligned with $h$ is a random variable even in a fully
  antiferromagnetically ordered configuration.} A canonical
simulation cannot separate the different staggered magnetization sectors, so a
study in this statistical ensemble will always be contaminated by this
spurious metastability and dominated by extremely rare events.

\subsection{The DAFF in the tethered ensemble}\label{sec:daff-th}
Following the results of our canonical investigation, the obvious quantity to tether
in the DAFF is the staggered magnetization $m_\text{s}$. 
This would, for instance, allow us to study separately the different saddle 
points, as discussed in Section~\ref{sec:ensemble-equivalence}, and, through the Helmholtz potential,
to determine their relative weight, without any need for exponentially slow tunneling.
In addition, since we are interested in a transition at non-zero magnetic field $h$,
in which metastability could also appear, 
we will tether the standard magnetization $m$. 

We, therefore, have a two-component tethered magnetic field
\begin{equation}\label{eq:DAFF-Omega}
\boldsymbol\nabla \Omega_N(\hat m,\hat m_\text{s}) = \biggl( \frac{\partial \Omega_N(\hat m,\hat m_\text{s})}
{\partial \hat m},\ \frac{\partial \Omega_N(\hat m,\hat m_\text{s})}{\partial \hat m_\text{s}}\biggr)
= \bigl( \langle\hat b\rangle_{\hat m,\hat m_\text{s}},\langle \hat b_s\rangle_{\hat m,\hat m_\text{s}}\bigr)
\end{equation}

We use quadratically added demons, as in Eq.~\eqref{eq:demonios-cuadraticos}.
Therefore,
\begin{align}\label{eq:DAFF-tethered-magnetic-field}
\hat b &=  1-\frac{1/2-1/N}{\hat m-m},\\
\hat b_\text{s} &=  1-\frac{1/2-1/N}{\hat m_\text{s}-m_\text{s}}.
\end{align}

In general, canonical averages in the presence of an external magnetic 
field with a regular component $h$ and a staggered component $h_\text{s}$
can be recovered through the Legendre transform equation:
\begin{equation}\label{eq:DAFF-tethered-to-canonical}
\langle O\rangle(h,h_\text{s}) = \frac{\int\dd \hat m\int \dd \hat m_\text{s}\ \langle O\rangle_{\hat m,\hat m_\text{s}}\
\ee^{-N [\Omega_N(\hat m,\hat m_\text{s})-\hat m \beta h-\hat m_\text{s} \beta h_\text{s}]}}
{\int\dd \hat m\int \dd \hat m_\text{s}\ 
\ee^{-N [\Omega_N(\hat m,\hat m_\text{s})-\hat m \beta h-\hat m_\text{s} \beta h_\text{s}]}},
\end{equation}
although here we shall be interested in $h_\text{s}=0$, so we shall use
the notation
\begin{equation}
\langle O\rangle(h) = \langle O\rangle (h,h_\text{s}=0).
\end{equation}

\begin{figure}
\centering
\includegraphics[height=.75\columnwidth,angle=270]{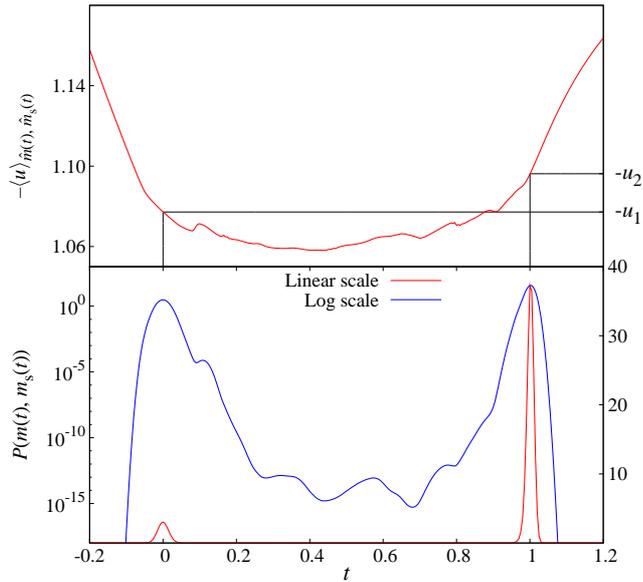}
\caption{(Color online) \emph{Top:} Result of a tethered simulation
  for the sample of Figure~\ref{fig:historia_canonico}. The two saddle
  points are joined by a straight path $(\hat m(t),\hat
  m_\text{s}(t))$, along which we measure the energy density $\langle
  u\rangle_{\hat m(t),\hat m_\text{s}(t)}$.  For $t=0$ and $t=1$ we
  reproduce the energies of the local minima obtained in the canonical
  simulation. \emph{Bottom:} Relative weight of the points along the
  simulated path, obtained from the line integral of the tethered
  magnetic field~\eqref{eq:DAFF-tethered-magnetic-field} (see text).}
\label{fig:segmento}
\end{figure}

The computation of these canonical averages following
the procedure described in Section~\ref{sec:tmc}
is very complicated. It would involve the non-trivial
computation of a two-dimensional potential from its gradient, after running
a set of many simulations in a two-dimensional grid in $(\hat m,\hat m_\text{s})$.
As we shall see in the following, the practical application 
is simpler than that. This is because for a given value of $h$ only a 
very small region of the configuration space $(\hat m,\hat m_\text{s})$ will have
a relevant weight in~(\ref{eq:DAFF-tethered-to-canonical}),
as discussed in Section~\ref{sec:ensemble-equivalence}. Therefore, due
to the ensemble equivalence property, we will be able to relate 
canonical and tethered averages through saddle-point equations
\begin{equation}\label{eq:DAFF-saddle-point-sample}
\begin{cases}
\displaystyle
\frac{\partial \Omega_N}{\partial \hat m} &= \langle\hat b\rangle_{\hat m,\hat m_\mathrm{s}} = \beta h,\\ 
\\
\displaystyle
\frac{\partial \Omega_N}{\partial \hat m_\mathrm{s}} &= \langle\hat b_\mathrm{s}\rangle_{\hat m,\hat m_\mathrm{s}}  = 0. 
\end{cases}
\end{equation}

As an example we can return to the sample of Section~\ref{sec:DAFF-canonical}. 
Using a canonical parallel-tempering simulation, we had identified 
two local minima, but were unable to determine their relative weights.
We can perform now this study in the tethered formalism.

The first step in this computation is identifying the values
of $(\hat m,\hat m_\text{s})$ that correspond to the local minima
 using~\eqref{eq:DAFF-saddle-point-sample}.
Notice that from the canonical simulation we know the values of $m^{(i)}$ and
$m_\text{s}^{(i)}$, so, using Eq.~(\ref{eq:DAFF-tethered-magnetic-field}),
$\hat m_\text{s}^{(i)} \simeq m_\text{s}^{(i)}+1/2$
and $\hat m^{(i)} \simeq m^{(i)} + 1/(2(1-\beta h))$. From
these starting guesses the actual minima are readily found.
The strategy is then clear: by performing the line integral of
\begin{equation}
\hat {\boldsymbol B}=\bigl(\langle \hat b\rangle_{\hat m,\hat m_\text{s}} - \beta h, \ \ 
\langle\hat b_\mathrm{s}\rangle_{\hat m,\hat m_\text{s}}\bigr)
\end{equation}
along a path connecting the two saddle
points we immediately obtain their potential difference or, 
equivalently, the ratio between their weights. The very same procedure,
carried out for hard-spheres, is depicted in Fig.~\ref{fig:grid}.

Figure~\ref{fig:segmento}
shows the result of this  computation. For our connecting path we use the straight line
$(\hat m(t), \hat m_\text{s}(t)) = (\hat m^{(1)}, \hat m_\text{s}^{(1)}) (1-t) + (\hat m^{(2)} ,
\hat m_\text{s}^{(2)})$. As we can see in the upper panel,
the tethered averages $\langle u\rangle_{\hat m(t), \hat m_\text{s}(t)}$ 
for $t=0$ and $t=1$ closely match $u^{(1)}$ and $u^{(2)}$, confirming
ensemble equivalence. The lower panel
shows the relative weight $P(\hat m(t),\hat m_\text{s}(t))$
of the points along the path. In accordance with \eqref{eq:DAFF-tethered-to-canonical}
this is simply
\begin{equation}
P(\hat m(t) ,\hat m_\text{s}(t)) = \ee^{-N \bigl[\Omega_N (\hat m(t), \hat m_\text{s}(t))- \beta h \hat m(t)\bigr]},
\end{equation}
where we have chosen the zero in $\Omega_N$ so that the  
weight of the whole path is normalized.

One of the two peaks is seen to have 
about ten times more importance than the other. Interestingly, we see 
that they are separated by a region of very low probability, which explains
the difficulty of the canonical simulations to tunnel between the two of them.
The large number of intermediate maxima and minima in the effective potential
are a real effect, and not fluctuations, as we shall see more clearly in the next section.

\subsection{Self-averaging and the disorder average}\label{sec:DAFF-disorder-average}
In order to perform a quantitative analysis of the DAFF, we have to
simulate a large number of samples and perform the disorder average.
The naive way of doing this for a system
with quenched disorder would be to measure $\hat{\boldsymbol B}$
and construct $\Omega_N$ for each sample,
then use Eq.~(\ref{eq:DAFF-tethered-to-canonical}) 
to compute all the  physically relevant $\langle O \rangle(h)$. 
Only then would we average $\langle O\rangle(h)$ over the disorder.

This approach is, however, paved with pitfalls. First of all, computing 
a two-variable $\Omega_N(\hat m,\hat m_\text{s})$ from a two-dimensional 
$(\hat m,\hat m_\text{s})$ grid is not an easy matter. In the previous section
we avoided this problem by finding the local minima first and then
evaluating $\Omega_N$ only along a path joining them. But this cannot be done 
efficiently and safely for a large number of samples. Even if it could be done,
the free-energy landscape of each sample is very complicated, with many local 
minima, several of which could be relevant to the problem, so a very high
resolution would be needed on the simulation grid. 

Finally, even reliably and efficiently computing 
the canonical averages $\langle O\rangle(h)$ 
would not be the end of our problems. Indeed, 
it is a well known fact that random systems often suffer from
violations of self-averaging~\cite{aharony:96,wiseman:98,malakis:06}.
This phenomenon has recently been studied in detail for 
the RFIM~\cite{parisi:02}, where the violations of
self-averaging are shown to be especially severe.
In this section we address
all these problems and demonstrate the computational strategy
followed in our study of the DAFF.

The first step is ascertaining whether the tethered averages themselves
are self-averaging or not. We already know that, since we are going to 
be working with a large regular external field $h$, the relevant 
region for the regular magnetization $m$ (and hence $\hat m$)
is going to be very narrow. Therefore, we are going to explore
the whole range of $\hat m_\text{s}$ for a fixed value of $\hat m=0.12$.

\begin{figure}
\centering
\begin{minipage}{.495\linewidth}
\includegraphics[height=1.08\columnwidth,angle=270]{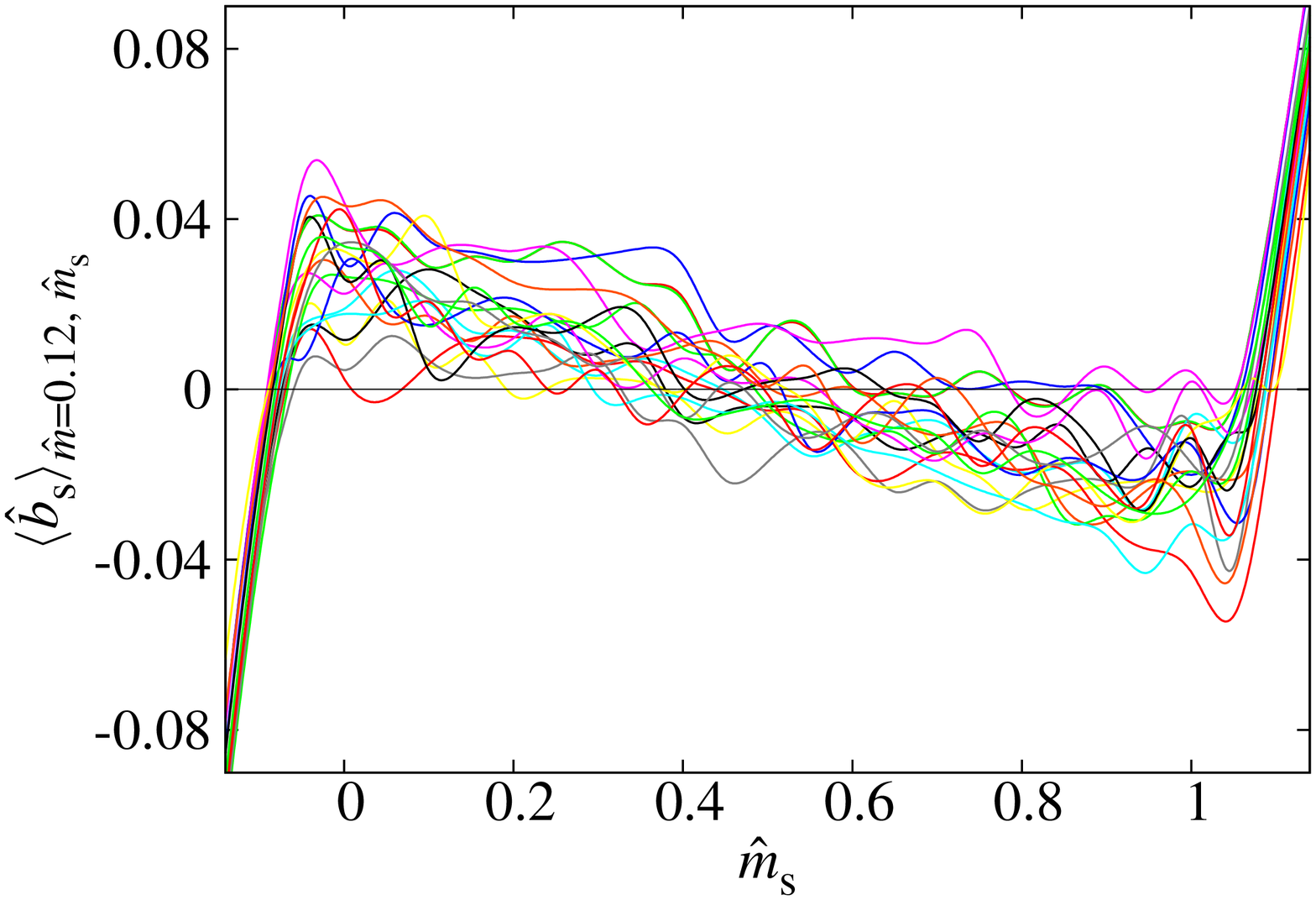}
\end{minipage}
\begin{minipage}{.495\linewidth}
\includegraphics[height=1.08\columnwidth,angle=270]{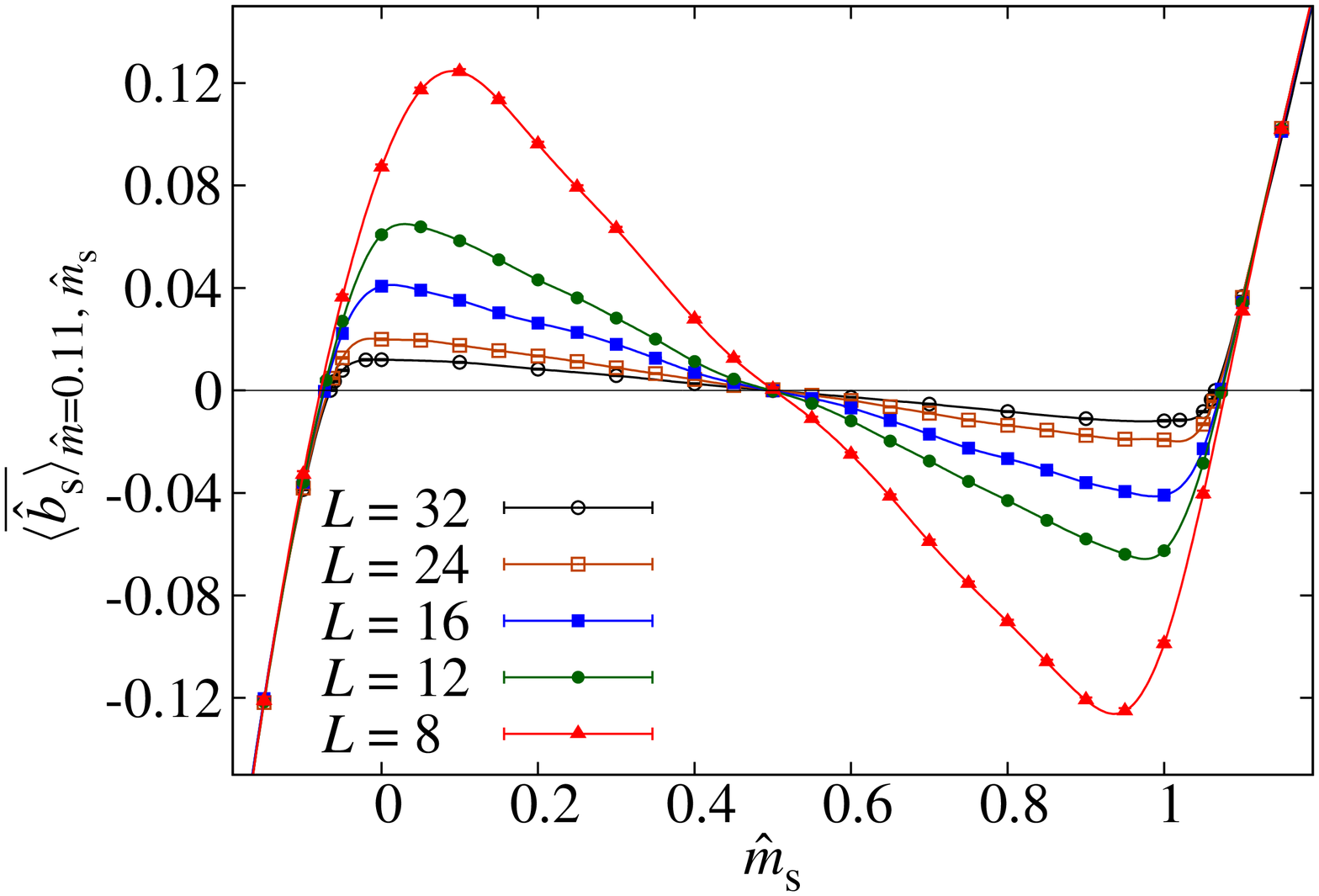}
\end{minipage}
\caption{(Color online) \emph{Left:} Staggered component of the
  tethered magnetic field, $\langle \hat b\rangle_{\hat m,\hat
    m_\text{s}}$ for $\hat m=0.12$ as a function of $\hat
  m_\text{s}$. We plot the results for several samples of an $L=24$
  system at $\beta=0.625$.  The curves are cubic splines interpolated
  from $33$ simulated points.  \emph{Right:} Disorder-averaged
  $\overline{\langle \hat b_\text{s}\rangle}_{\hat m,\hat m_\text{s}}$
  for the same smooth magnetization and temperature as the left panel,
  for all our system sizes. The plot shows the different behavior of
  the regions inside the gap, where the staggered magnetic field goes
  to zero as $L$ increases, and outside of it, where there is a
  non-zero enveloping curve.}
\label{fig:hatbs-samples}
\end{figure}
We have plotted the staggered tethered magnetic field
 $\langle \hat b_\text{s}\rangle_{\hat m=0.12,\hat m_\text{s}}$
for $20$ samples of an $L=24$ system at $\beta=0.625$
in Figure~\ref{fig:hatbs-samples}---\emph{left}.  The different 
curves have a variable number of zeros, but all of them have at least three:
one in the central region and two roughly symmetrical ones for large 
staggered magnetization.  The positions
of the two outermost zeros  clearly separate two differently behaved 
regions. Inside of the gap the sample-to-sample fluctuations are chaotic,
while outside of it the sheaf of curves even seems to have an envelope.
This impression is confirmed in the right panel of Figure~\ref{fig:hatbs-samples},
where we show the sample-averaged tethered magnetic field for several system
sizes.

\begin{figure}
\centering
\begin{minipage}{.49\linewidth}
\includegraphics[height=1.05\columnwidth,angle=270]{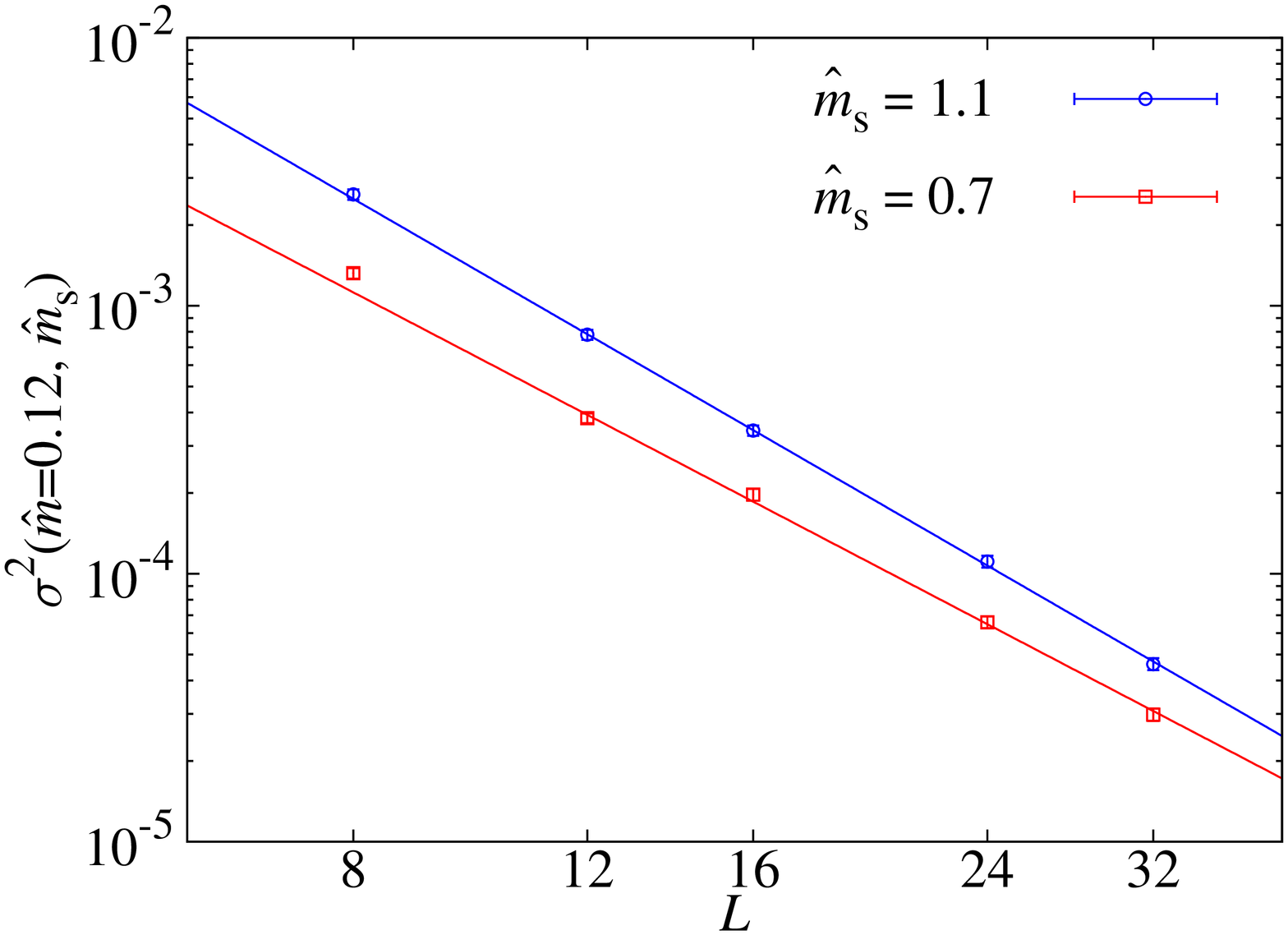}
\end{minipage}
\begin{minipage}{.49\linewidth}
\includegraphics[height=1.05\columnwidth,angle=270]{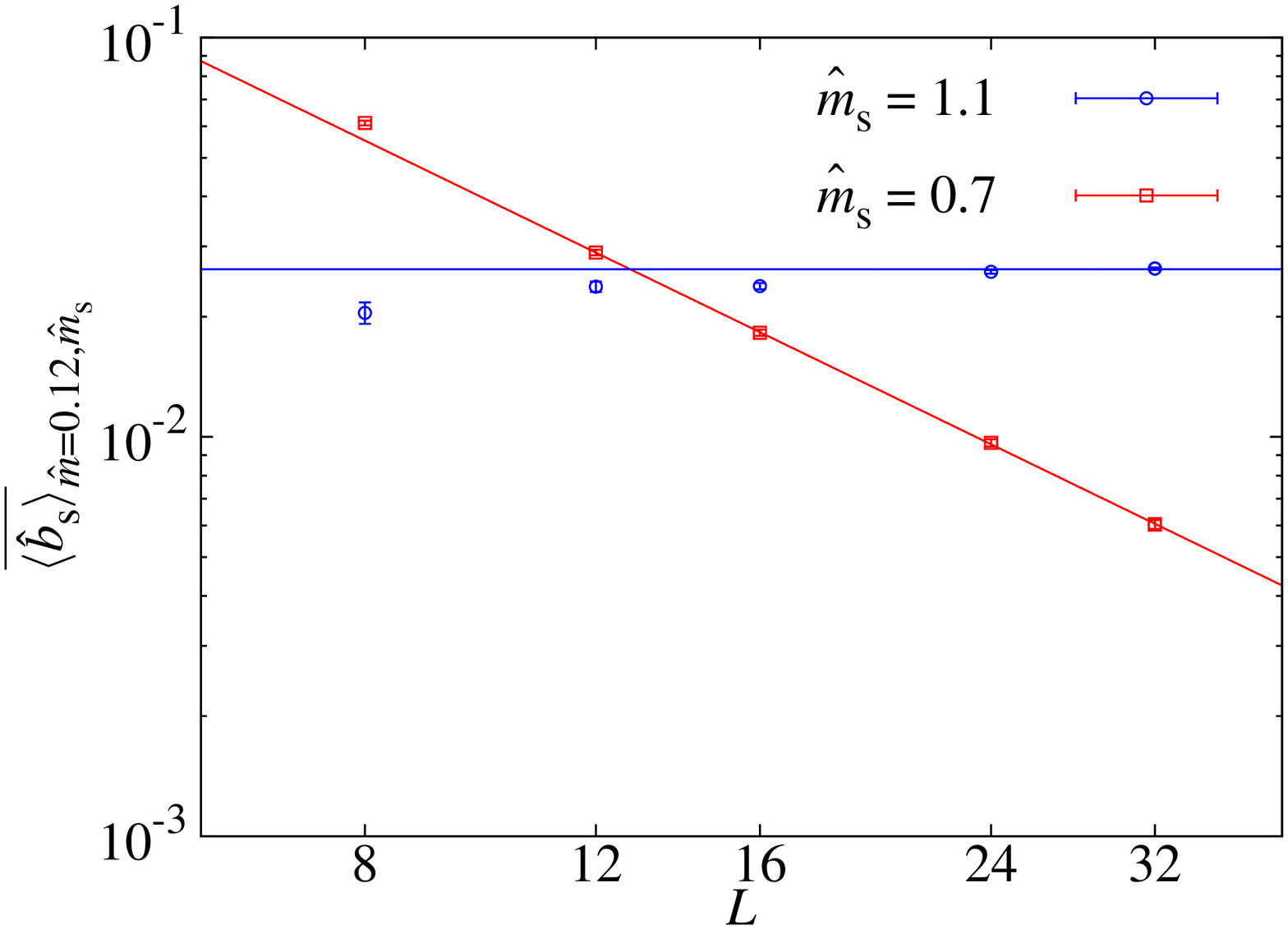}
\end{minipage}
\caption{(Color online) \emph{Left:} Sample-to-sample fluctuations in
  the staggered component of the tethered magnetic field,
  Eq.~\eqref{eq:DAFF-sigma}, against system size. We show two sets of
  points, both at $\beta=0.625$ and $\hat m=0.12$, but at different
  values of $\hat m_\text{s}$.  The red curve, corresponding to $\hat
  m_\text{s}=0.7$ is right inside the gap defined by the outermost
  zeros in Figure~\ref{fig:hatbs-samples}, while the blue curve ($\hat
  m_\text{s}=1.1$) is outside. Both are shown to decay with a power
  law.  \emph{Right:} Disorder average $\overline{\langle\hat
    b_\text{s}\rangle}_{\hat m,\hat m_\text{s}}$ for the systems of
  the left panel. Inside the gap, the field goes to zero with $L$;
  outside it has a finite limit.}
\label{fig:automediancia} 
\end{figure}
In order to quantify this  observation we can study the fluctuations of the 
disorder-averaged $\overline{\langle \hat b_\text{s}\rangle}_{\hat m,\hat m_\text{s}}$,
\begin{eqnarray}
\sigma^2(\hat m,\hat m_\mathrm{s}) &=&  \overline{\left({\langle\hat b_\mathrm{s}\rangle}_{\hat m,\hat m_\mathrm{s}} - \overline{\langle\hat b_\mathrm{s}\rangle}_{\hat m,\hat m_\mathrm{s}} \right)^2}\ \label{eq:DAFF-sigma}.
\end{eqnarray} 
This quantity is plotted on the left panel of Figure~\ref{fig:automediancia}. As we
can see, it goes to zero as a power in $L$, so we also plot fits to
\begin{equation}
\sigma^2(\hat m=0.12,\hat m_\text{s}) = A(\hat m_\text{s}) L^{c(\hat m_\text{s})}.
\end{equation} 
This would seem like a very good
sign, because it could be indicative of self-averaging behavior. However,
it is not the whole story. If we recall the right panel
of Figure~\ref{fig:hatbs-samples}, we see that inside the gap the tethered magnetic
field itself, not only its fluctuations, goes to zero as $L$ increases. In fact, as shown 
in Figure~\ref{fig:automediancia}, for
$\hat m_\text{s} = 0.7$,  $\sigma^2\sim L^{-2.5}$, while 
$\overline{\langle \hat b_\text{s}\rangle}_{\hat m,\hat m_\text{s}}\sim L^{-1.6}$.
This means that the relative fluctuations $\sigma/\overline{\langle \hat b_\text{s}\rangle}_{\hat m,\hat m_\text{s}}$ do not decrease with increasing $L$.
For the point outside the gap, however, the disorder average of the tethered magnetic 
field reaches a plateau. Furthermore, the fluctuations decay with $c=2.94(7)$, which
is compatible with the value $c=D$ that one would expect in a self-averaging system.

In physical terms, this analysis means that the local minima for a small, but non-zero,
value of the applied staggered magnetic field $h_\text{s}$ would be self-averaging.
We can now recall the well-known recipe for dealing with spontaneous 
symmetry breaking: consider a small applied field and take the thermodynamical
limit \emph{before} making the field go to zero.
Translated to the DAFF, this means that we should solve the saddle-point
equations~\eqref{eq:DAFF-saddle-point-sample}
on average, rather than sample by sample, and then take the $h_\text{s}\to0$
limit on the results,
\begin{equation}\label{eq:saddle-point-medio}
\begin{cases}
\displaystyle
\frac{\partial \overline\Omega_N}{\partial \hat m} &= \overline{\langle\hat b\rangle}_{\hat m,\hat m_\mathrm{s}} = \beta h,\\ 
\\
\displaystyle
\frac{\partial \overline\Omega_N}{\partial \hat m_\mathrm{s}} &= \overline{\langle\hat b_\mathrm{s}\rangle}_{\hat m,\hat m_\mathrm{s}}  = 0^+. 
\end{cases}
\end{equation}
In other words, we are considering the disorder average of a thermodynamical
potential, $\Omega_N$, different from the free energy. This approach was
first introduced in~\cite{fernandez:08}, in a microcanonical context
(the averaged potential was in that case the entropy).

\begin{figure}
\centering
\includegraphics[height=.85\columnwidth,angle=270]{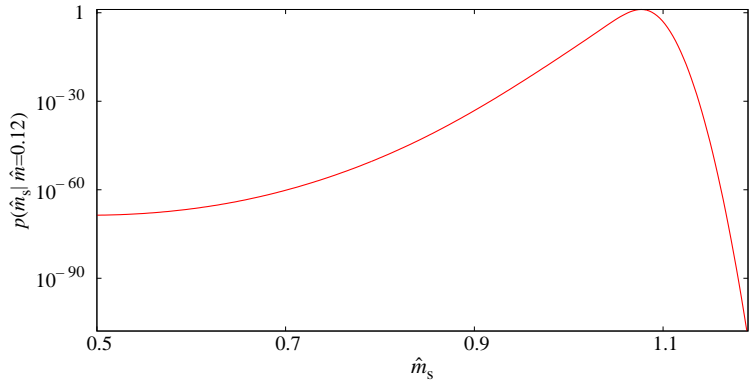}
\caption{(Color online) Disorder-averaged probability density function
  of $\hat m_\text{s}$ conditioned to $\hat m = 0.12$ for the system
  of Figure~\ref{fig:hatbs-samples}.}\label{fig:P-hatms}
\end{figure}

Notice that for the regular component of the tethered magnetic field
we are always going to work with a finite value of the external field $h$.
Therefore, the integral in~\eqref{eq:DAFF-tethered-to-canonical} is going to be completely
dominated by an extremely narrow $\hat m$ range. Therefore, we can consider
different values of $\hat m$ separately and relate them to the particular
$h$ that generates a local minimum at that magnetization.
This is best accomplished by integrating the disorder-averaged 
tethered magnetic field along 
the path $\bigl(\hat m=\hat m_0, \hat m_\text{s}(t)\bigr)$. In this
way we obtain the probability 
distribution of $\hat m_\text{s}$, conditioned to $\hat m =\hat m_0$, which we will
denote by $p(\hat m_\text{s} | \hat m_0)$ (Figure~\ref{fig:P-hatms}). This 
probability density function can be used to average over $\hat m_\text{s}$ for 
fixed $\hat m$,
\begin{equation}\label{eq:promedio-hatms}
\overline{\langle O\rangle}_{\hat m} = \int \dd \hat m_\text{s} \ p(\hat m_\text{s} | \hat m)
\overline{\langle O\rangle}_{\hat m,\hat m_\text{s}} =
\int \dd \hat m_\text{s} \ \overline{\langle O\rangle}_{\hat m,\hat m_\text{s}}  \ee^{-N \overline\Omega_N(\hat m_\text{s} | \hat m)},
\end{equation}
where 
\begin{equation}\label{eq:Omega1}
\frac{\partial \overline \Omega_N(\hat m_\text{s}|\hat m)}{\partial \hat m_\text{s}} = 
\overline{\langle \hat b_s\rangle}_{\hat m,\hat m_\text{s}}\ .
\end{equation}

With this process we have integrated out the dependence on $\hat m_\text{s}$.
We now have a series of smooth functions $\overline{\langle O\rangle}_{\hat m}$,
together with a smooth one-to-one function
$h(\hat m) = \overline{\langle \hat b\rangle}_{\hat m}/\beta$. Remembering
our discussion of ensemble equivalence in Section~\ref{sec:ensemble-equivalence},
we shall make the approximation
\begin{equation}
\overline{ \langle O\rangle}(h) \simeq \overline{\langle O\rangle}_{\hat m(h)}.
\end{equation}
At any rate, we remark that the r.h.s. is better suited for reproducing the
physics of experimental samples.

Finally, let us mention that some of the physical observables
are correlated with the distribution of the $\epsilon_\bx$.
This can be exploited to reduce the statistical errors
of their sample averages, using the technique of control
variates~\cite{fernandez:09c,yllanes:11}.

\subsection{Our simulations}\label{sec:DAFF-simulations}
\begin{table}
\centering
\caption{Parameters of our simulations. For each of the $N_\mathrm{samples}$
disorder realizations for each $L$,
 we run $N_{\hat m} \times N_{\hat m_\mathrm{s}}$
tethered simulations with temperature parallel tempering.
 The $N_T$ participating temperatures
are evenly spaced in the interval $[1.6,2.575]$. For each size
we report the minimum number of Monte Carlo steps
 (Metropolis sweep + parallel tempering). After the application of
our thermalization criteria, some of the simulations for $L\geq24$ 
needed to be extended, leading to a higher average number of
Monte Carlo steps ($N_\mathrm{MC}^\mathrm{av}$).}
\begin{tabular*}{\columnwidth}{@{\extracolsep{\fill}}rcccccc}
\hline
 $L$ & $N_\mathrm{samples}$ &$N_T$ &  $N_{\hat m}$ &
 $N_{\hat m_\mathrm{s}} $ & $N_\mathrm{MC}^{\mathrm{min}}$ 
& $N_\mathrm{MC}^\mathrm{av}$ \\
\hline
8&  1000 & 20 & 5 &  31 & $7.7\times10^5$ & $7.7\times10^5$  \\
12& 1000 & 20 & 5 &  35 & $7.7\times10^5$ & $7.7\times10^5$  \\
16& 1000 & 20 & 5 &  35 & $7.7\times10^5$ & $7.7\times10^5$  \\
24& 1000 & 40 & 5 &  33 & $7.7\times10^5$ & $9.3\times10^5$  \\
32&  700 & 40 & 4 &  25 & $1.5\times10^6$ & $5.5\times10^6$  \\
\hline
\end{tabular*}
\label{tab:parametros-PT}
\end{table}
We can infer several useful conclusions from the analysis of the previous section 
\begin{itemize}
\item The disorder average should be performed on the tethered
observables, before computing the effective potential.
\item It is best to analyze several values of $\hat m$ separately,
since the average over $\hat m_\text{s}$ for each fixed $\hat m$ can be 
unambiguously related to the canonical average via
$h(\hat m) = \overline{\langle \hat b\rangle}_{\hat m} / \beta$.
In this way, we can study the phase transition that arises by varying 
the applied magnetic field at fixed $\beta$.
\item For fixed $\hat m$ the conditioned probability $p(\hat m_\text{s} | \hat m)$ 
has two narrow, symmetric peaks, separated by a region with extremely 
low probability.
\end{itemize}
Therefore, we have carried out the following steps
\begin{enumerate}
\item Select an appropriate grid of $\hat m$ values. This
should be wide enough to include the critical point for the simulation
temperature, and fine enough to detect the fluctuations of
$\overline{\langle O\rangle}_{\hat m}$. These turn out
to be very smooth functions of $\hat m$, so a few values of this parameter
suffice~\cite{fernandez:11b}.
\item For each value of $\hat m$, select an appropriate grid of $\hat m_\text{s}$.
We start with evenly spaced points and after a first analysis add more
values of $\hat m_\text{s}$ in the neighborhood of the minima, as this
is the more delicate and relevant region.
\item The simulations for each $(\hat m,\hat m_\text{s})$ are carried out 
with the Metropolis update scheme of section~\ref{sec:Metropolis}.
In addition, we use parallel tempering (attempting to exchange configurations
at neighboring temperatures). This is not needed in order to thermalize the system
for $L<32$, but it is convenient since we also study
the temperature dependence of some observables.
Furthermore, the use of parallel-tempering provides a reliable 
thermalization check (see the Appendix, particularly section~\ref{sec:thermalization-PT}).
\end{enumerate}
From Figure~\ref{fig:P-hatms} it seems that, in addition, it would pay to
restrict the simulations to a narrow range of $\hat m_\text{s}$ around the
peaks, since the tethered values for $\hat m_\text{s}$ away from them cannot
possibly contribute to the average. This is true but for one exception, the
computation of the hyperscaling violation exponent~$\theta$, which requires
that the whole range be explored (see~\cite{fernandez:11b}).\footnote{In
  Section~\ref{sec:DAFF-geometry} we show an additional study that requires
  simulations far from the peaks.}

The parameters of our simulations are presented on Table~\ref{tab:parametros-PT}. 
The table lists the number $N_{\hat m}$ of values in our $\hat m$ grid
and the number of points in the $\hat m_\mathrm{s}$ grid for each, so 
the total number of tethered simulations for each sample is $N_{\hat m}\times N_{\hat m_\mathrm{s}}$.

The number of Monte Carlo steps in each tethered simulation 
is adapted to the autocorrelation time (see next section), the table
lists the minimum length $N_\text{MC}^\text{min}$ 
and the average length $N_\text{MC}^\text{av}$ for each lattice size.

\subsection{Thermalization and metastability in the tethered simulations}\label{sec:DAFF-thermalization}
We have assessed the thermalization of our simulations through the 
autocorrelation times of the system, computed from the analysis
of the parallel tempering dynamics (see Section~\ref{sec:thermalization-PT}).
We  require a simulation time longer than
$100\tau_\mathrm{int}$.\footnote{For most 
simulations $\tau_\text{int} \simeq \tau_\text{exp}$, so
this value is much larger than what is required 
to achieve thermalization. This ample choice of minimum
simulation time protects us from the few cases where $\tau_\text{exp}$
is noticeably larger than $\tau_\text{int}$.}
As a fail-safe mechanism, we compute the time $t_\mathrm{hot}$ that 
each configuration spends in the upper half of the temperature range.
If any of these is less than a third of the median $t_\mathrm{hot}$, we 
double the simulation time, considering that the simulation is too
short to allow a good determination of $\tau_\mathrm{int}$. 
This process is only followed for $L\geq24$. For smaller
sizes we have simply made the minimum simulation 
time large enough to thermalize all samples.

\begin{figure} \centering
\includegraphics[height=0.7\columnwidth,angle=270]{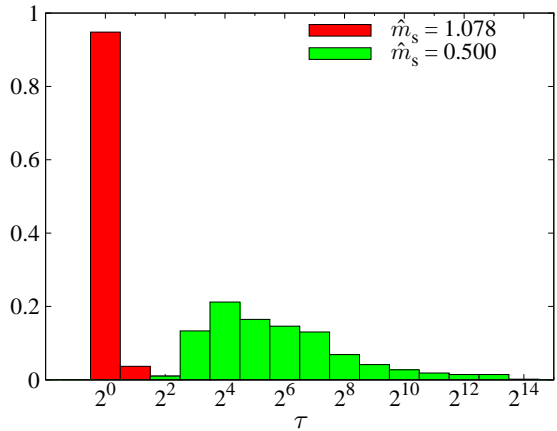}
\caption{(Color online) Histograms of thermalization times for our $L=32$ simulations,
at two values of $\hat m_\mathrm{s}$ for $\hat m=0.12$. Notice 
the logarithmic scale in the horizontal axis, which is in units 
of $300$ parallel-tempering steps. The minimum of the effective potential
(the peak in the probability distribution) at this $\hat m=0.12$
corresponds to $\hat m_\mathrm{s}=1.078$. Notice 
that we cannot measure times shorter than our measuring
frequency of $300$ parallel-tempering steps, so the first 
bin should be taken as encompassing all shorter autocorrelation
times. Only the samples with $\tau\gtrsim 2^6$ have to 
be extended from the minimum simulation time.}
\label{fig:histograma_hatm_012}
\end{figure}

The distribution of correlation times for our different samples
turns out to be dependent on the value of $(\hat m,\hat m_\mathrm{s})$.
Considering first the variation of the average $\tau_\mathrm{int}$
with $\hat m_\mathrm{s}$ at fixed $\hat m$, we see that the
minimum and its adjoining region 
are much easier to thermalize~(Figure~\ref{fig:histograma_hatm_012}).
This region coincides with the only points that have a non-negligible
probability density~(Figure~\ref{fig:P-hatms}), i.e., the only points 
that contribute to the computation of the $\overline{\langle O\rangle}_{\hat m}$.
This fact suggests a possible optimization, that we will discuss in 
Section~\ref{sec:DAFF-optimization}.

\begin{figure} \centering
\includegraphics[height=0.7\columnwidth,angle=270]{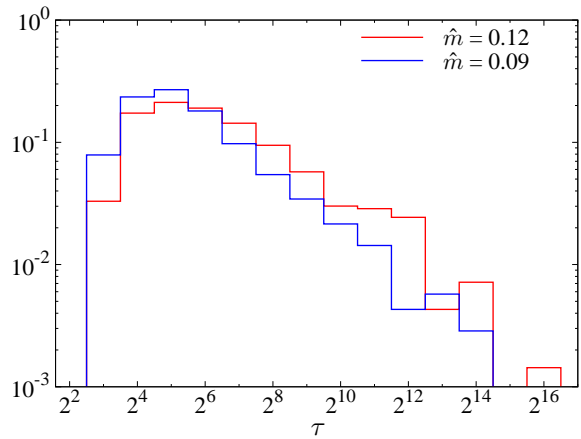}
\caption{(Color online) Same as Fig.~\ref{fig:histograma_hatm_012}, but now we consider
  several values of $\hat m$ for $\hat m_\mathrm{s}=0.8$ (in the hard
  thermalization region far from the peak). The points for $\hat m=0.12$,
  closer to the critical point, exhibit a much heavier long-times tail (mind
  the vertical logarithmic scale).}
\label{fig:histograma_hatms_08}
\end{figure}

A second interesting result comes from studying the evolution of the $\tau_{\mathrm int}$ 
with $\hat m$. Figure~\ref{fig:histograma_hatms_08} represents the histogram
of autocorrelation times for $\hat m_\mathrm{s}=0.8$ (in the `hard' region)
for two values of $\hat m$. The distribution for $\hat m=0.12$ has a much 
heavier tail. It is shown in~\cite{fernandez:11b} that this is due
to the onset of a phase transition. 

\begin{figure} \centering
\includegraphics[height=1.05\columnwidth,angle=270]{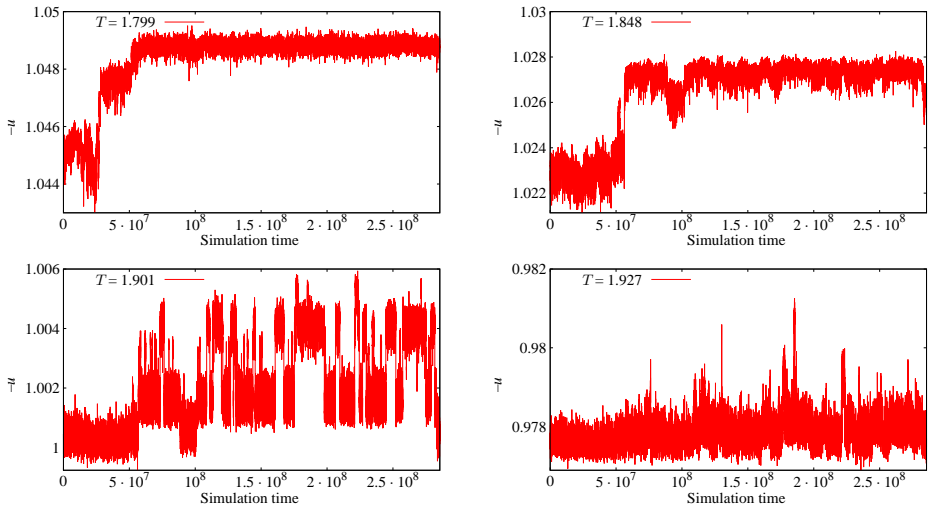}
\caption{(Color online) Time evolution (in number of parallel tempering steps) for the energy density $u$ 
for several temperatures of a single $L=32$ sample at $\hat m=0.12$, $\hat m_\text{s}=0.7$.
For most temperatures the equilibrium value is quickly reached, but
for a very narrow temperature range there are several competing 
metastable states (bottom-left panel for $T=1.901$).
}
\label{fig:metaestabilidad}
\end{figure}

\begin{figure}
\centering
\includegraphics[height=.85\columnwidth,angle=270]{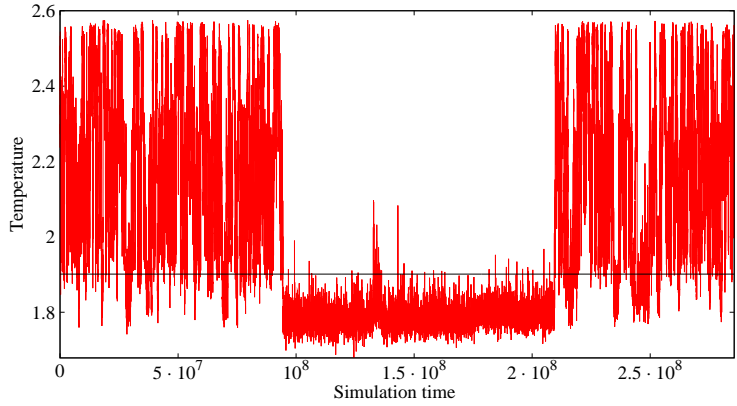}
\caption{(Color online) Temperature random walk for one of the 40 replicated systems for the
  parallel tempering simulation of the sample in
  Figure~\ref{fig:metaestabilidad}.  The flow is blocked at the same
  temperature that had several metastable states (marked with a horizontal
  line), see bottom-left panel in Fig.~\ref{fig:metaestabilidad}.}
\label{fig:historia_metaestabilidad}
\end{figure}

The difficulty in thermalizing some samples stems from the coexistence
of several metastable states, even for fixed $(\hat m,\hat m_\mathrm{s})$.
In Figure~\ref{fig:metaestabilidad} we represent the time evolution of the 
energy $u$ for several temperatures of the same sample ($L=32$, $\hat m=0.12$, 
$\hat m_\text{s}=0.8$). As we can see, for a narrow temperature range several
metastable states compete. This has a very damaging effect on the parallel
tempering dynamics, that get stuck whenever a configuration that is metastable
for one temperature is very improbable in the next (see Figure~\ref{fig:historia_metaestabilidad}).

\begin{figure} \centering
\begin{minipage}{.49\linewidth}
\includegraphics[height=1.05\columnwidth,angle=270]{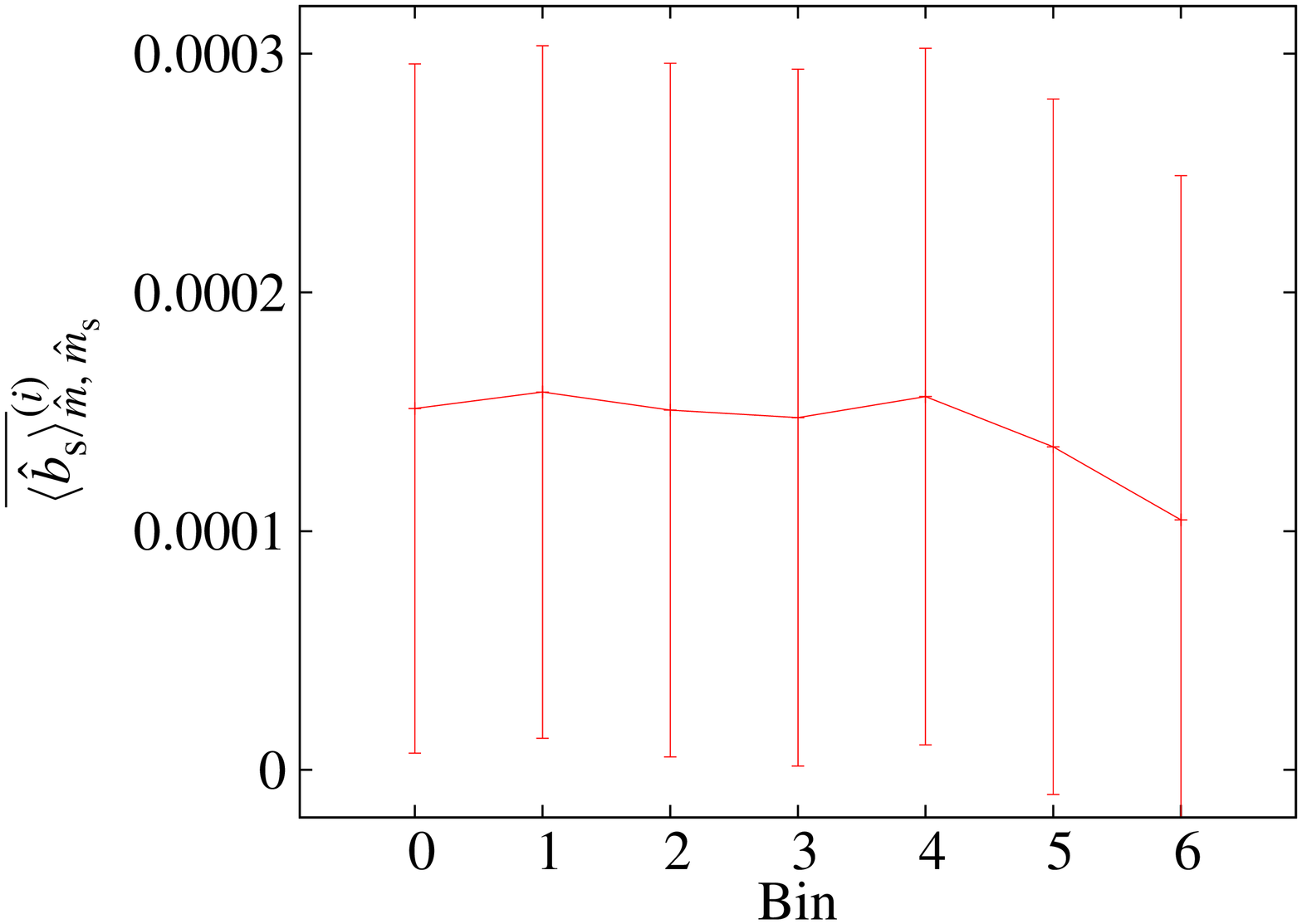}
\end{minipage}
\begin{minipage}{.49\linewidth}
\includegraphics[height=1.05\columnwidth,angle=270]{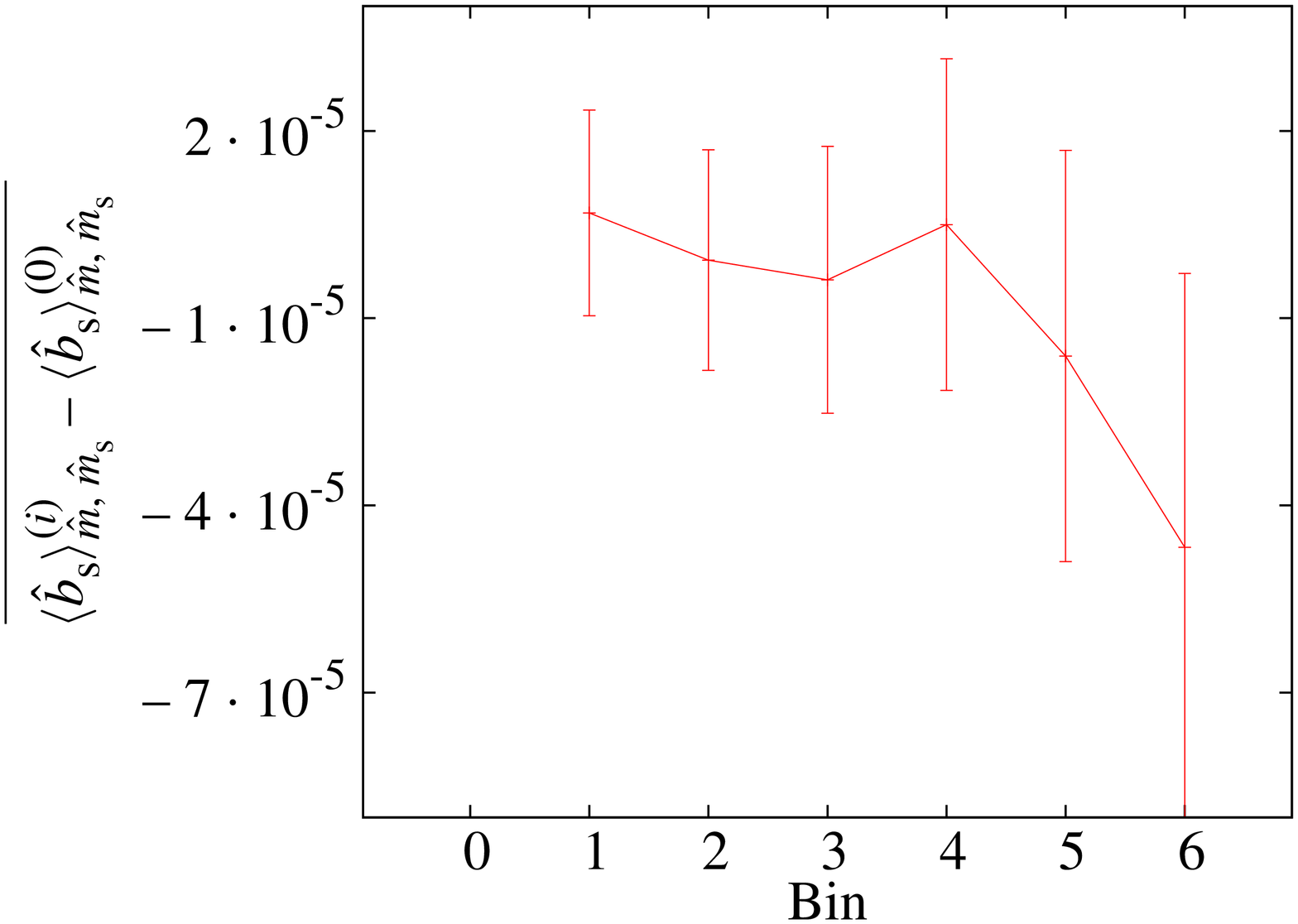}
\end{minipage}
\caption{(Color online) \emph{Left:} Time evolution of
  $\overline{\langle\hat b_\mathrm{s}\rangle}_{\hat m,\hat
    m_\mathrm{s}}$ for $\hat m=0.12$, $\hat m_\mathrm{s}=0.5$, in a
  logarithmic scale (the first bin is the average over the second half
  of the simulation, the second bin the average over the second
  quarter and so on). This is in the middle of the $(\hat m,\hat
  m_\mathrm{s})$ region where thermalization is hardest.
  \emph{Right:} Same figure, but now subtracting the value of the
  first bin from each of the following ones. Since the data are
  correlated, this subtraction causes a very significant error
  reduction, but the difference is still compatible with zero for
  several bins.}
\label{fig:log2}
\end{figure}

For $L=32$, some points\footnote{By `point' we mean
any of the $N_\mathrm{samples}\times N_{\hat m}\times N_{\hat m_\mathrm{s}}$
individual tethered simulations for each $L$.} presented a metastability so severe
that enforcing a simulation time longer than $100\tau_\mathrm{int}$ 
would require a simulation of more than $10^9$ parallel-tempering 
updates (one thousand times longer than 
our minimum simulation of $\sim 10^6$ steps). Thermalizing these points 
(which constitute about $0.1\%$ of the total) would have thus required
some $10^6$ extra CPU hours, with a wall clock of many months.
We considered this to be disproportionate to their physical
relevance (they are all restricted to a region far from the peaks  where the probability density
is $<10^{-40}$, see Figure~\ref{fig:P-hatms}).
Therefore, we have stopped these simulations at about $10\tau_\mathrm{int}$. 
This is still a more demanding thermalization criterion than 
is usual for disordered systems and
does not introduce any measurable bias in the physically relevant 
disorder-averaged observables. This can 
be checked in several ways:
\begin{itemize}
\item First of all, as we have already discussed,
these points are restricted to a region
in $\hat m_\mathrm{s}$ with probability density of at most $10^{-40}$.
Therefore, even if there were a bias it would not have any effect 
in the computation of canonical averages.
\item For the only affected physical observable, the free-energy
barriers used to compute $\theta$ (see~\cite{fernandez:11b}),
we can compare the result 
for $L=32$ with the extrapolation from smaller sizes, and it is compatible.
\item Even at the most difficult values of $(\hat m,\hat
  m_\mathrm{s})$ the log$_2$-binning plot (the only thermalization
  test typically used in disordered-systems simulations) presents many
  logarithmic bins of stability (Figure~\ref{fig:log2}---left).  Even
  if we subtract the result of the last bin from the others (in
  equilibrium, this substration should yield zero), taking into account
  statistical correlations~\cite{fernandez:07}, several bins of
  stability remain (Figure~\ref{fig:log2}---right). This is a very
  strict test, and one that even goes beyond physical relevance
  (because it reduces the errors dramatically from those given in the
  final results).
\end{itemize}

\subsubsection{Optimizing Tethered Monte Carlo simulations}\label{sec:DAFF-optimization}
We can highlight two interesting facts from the previous discussion
\begin{itemize}
\item Only a very narrow region around the minima of the effective potential has any 
significant weight for reconstructing canonical averages (Figure~\ref{fig:P-hatms}).
\item It is much harder to equilibrate the region far from the minima
(Figure~\ref{fig:histograma_hatm_012}).
\end{itemize}
Combining these two observations, it turns out that, if our only interest is
reconstructing canonical averages, we can achieve a qualitative improvement
in simulation time by simulating only a narrow range around the peak 
in $P(\hat m_\text{s}| \hat m)$ for each value of the smooth magnetization
$\hat m$. In fact, all the results discussed in~\cite{fernandez:11b} could 
have been computed in this simplified way, except for the hyperscaling 
violations exponent~$\theta$.

We have demonstrated this optimization by simulating $400$ samples 
of an $L=48$ system
for $\hat m=0.12$. We use only $N_{\hat m_\text{s}}=6$ (but we have
to increase the number of temperatures in the parallel-tempering 
to keep the exchange acceptance high). Table~\ref{tab:picos}
shows the value of $\hat m_\text{s}^\text{peak}$ as a function of $L$.

\begin{table}
\centering
\caption{Value of the peak position for our different system
sizes. The result for $L=48$ is of comparable accuracy, despite
being computed from only $6$ tethered simulations around 
the minimum of the effective potential. Notice that for $L\leq 32$ we can average over 
the positive and negative peaks, so the number of samples 
for these systems is effectively double the value shown 
in Table~\ref{tab:parametros-PT}.}
\begin{tabular*}{\columnwidth}{@{\extracolsep{\fill}}rrrc}
\hline
$L$ & $N_\text{samples}$ & $N_{\hat m_\text{s}}$ & $\hat m_\text{s}^\text{peak}-1/2$ \\
\hline
8  & $1000\times2$ & 31 & 0.58585(87) \\
12 & $1000\times2$ & 35 & 0.58239(54) \\
16 & $1000\times2$ & 35 & 0.58154(36) \\
24 & $1000\times2$ & 33 & 0.57839(24) \\
32 &  $700\times2$ & 25 & 0.57672(20) \\
48 &  400 &  6 & 0.57491(33) \\
\hline
\end{tabular*}
\label{tab:picos}
\end{table}

\subsection{Geometrical study of the critical configurations}\label{sec:DAFF-geometry}
\begin{figure}
\centering
\includegraphics[width=.7\columnwidth]{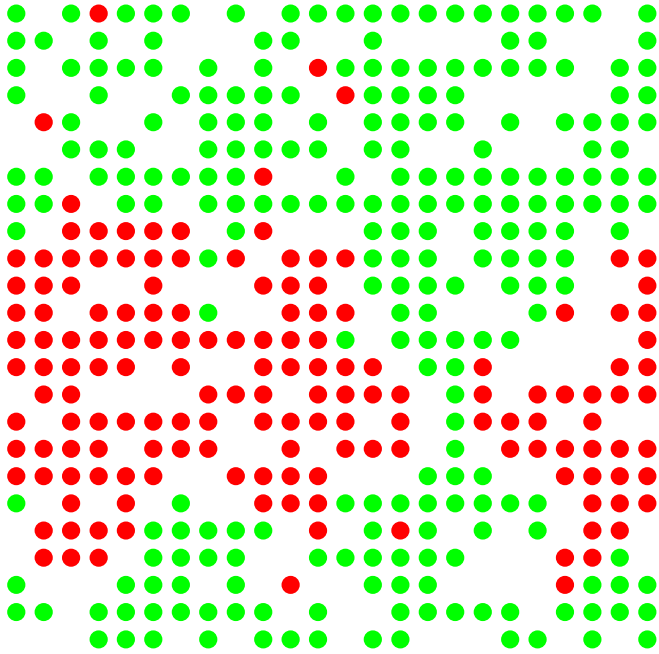}
\caption{(Color online) Equilibrium configuration for an $L=24$ system
at $\beta=0.625$, $\hat m=0.12$, $\hat m_\mathrm{s} = 0.5$.}
\label{fig:conf}
\end{figure}

The picture painted in~\cite{fernandez:11b} is that of a second-order
transition, but one with quite extreme behavior. Among its peculiarities
we can cite an extremely small value of $\beta$ and large free-energy
barriers, features both that are reminiscent of first-order behavior.
The free-energy barriers in the DAFF, however, diverge too slowly to be associated 
to the surface tension of well defined, stable coexisting phases. But
this only begs the question of what kind of configurations can give rise
to such a behavior.  

In this section we shall study the geometrical properties 
of the minimal-cost spin configurations joining the two 
ordered phases at the critical point. To this end, we 
consider simulations at $\beta=0.625$,
 $\hat m=0.12 \approx \hat m_\mathrm{critical}$ and $\hat m_\mathrm{s} = 0.5$.
Recalling that $\hat m_\mathrm{s} \simeq m_\mathrm{s} +1/2$, this 
last condition expresses the fact that we are studying configurations
with no global staggered magnetization. This is a good example of an 
`inherently tethered' study, that examines information hidden 
from a canonical treatment. 

Figure~\ref{fig:conf} shows an example of such a configuration
for an $L=24$ system.
In order to make the different phases clearer, we are not representing
the spin field $s_\bx$, but the staggered field $s_\bx\pi_\bx$. 
As is readily seen, even if the global magnetization is 
$m_\mathrm{s} \approx 0$, the system is divided into two phases
with opposite (staggered) spin. In geometrical terms, most of the 
occupied nodes of the system belong to one of two large clusters with 
opposite sign, with only a few scattered smaller clusters.  

At a first glance, this picture may seem consistent with a first-order
scenario, were the system is divided into two strips whose
surface tension gives rise to the free-energy barriers
(see~\cite{martin-mayor:07} for an example of these
geometrical transitions in a first-order setting). In order to test this
possibility, we can study the evolution of the interface mass with the
system size and compare it with the explicit computation of free-energy
barriers done in~\cite{fernandez:11b}.

Given a configuration, we first trace all the geometric 
antiferromagnetic clusters. We then identify the largest and second
largest ones. Finally, we say that an occupied node belongs
to the `interface' if it belongs to the largest cluster and has at least
one first neighbor belonging to the second largest one. We have 
computed in this way the interface $N_\mathrm{interface}$ for 
our $700$ $L=32$ samples and for $3000$ samples for all our
smaller systems (we have run additional simulations just at this point).
Table~\ref{tab:interfase} shows the result of this computation.
A fit to
\begin{equation}\label{eq:interfase}
N_\mathrm{interface} = A L^c,
\end{equation}
for $L\geq12$ gives $c=2.2402(24)$ with $\chi^2/$d.o.f. $=3.62/2$.  One may
think that such a large exponent could be indicative of a surface
tension. However, the study of~\cite{fernandez:11b} showed that free-energy
barriers in the DAFF grow as $L^\theta$, with $\theta=1.469(20)\neq
c$. Indeed, for the archetypical second-order model, the $D$-dimensional pure
Ising ferromagnet, the percolating clusters are space filling, so $c=D$.
\begin{table}
\centering
\caption{Masses of the interfaces for our equilibrium
spin configurations at $\beta=0.625$, $\hat m=0.12\approx\hat m_\mathrm{c}$,
$\hat m_\mathrm{s} =0 .5$. This mass grows as 
$N_\mathrm{interface} \propto L^c$. From a fit, $c=2.2402(24)$, which 
is incompatible with our estimate of $\theta=1.469(20)$, confirming
that the free-energy barriers are not associated to surface tensions.}
\begin{tabular*}{\columnwidth}{@{\extracolsep{\fill}}rrl}
\hline
 $L$ & $N_\mathrm{samples}$ & $N_\mathrm{interface}$\\
\hline
12 & 3000 & \phantom{1}160.29(25)\\
16 & 3000 & \phantom{1}304.62(41)\\
24 & 3000 & \phantom{1}755.74(94)\\
32 &  700 & 1446.1(39)\phantom{1}\\
\hline
\end{tabular*}
\label{tab:interfase}
\end{table}

A second interesting feature of the configuration pictured
on Figure~\ref{fig:conf} is that there is a path
connecting the spins in the green strip \emph{across} the red one 
(there is only one green strip, since we are considering periodic boundary conditions).
If we were to study a complete tomography of this configuration, 
we would find several  of these paths (which, of course, need not be
contained in a plane). In other words, the phases are porous. This is
in clear contrast to a first-order scenario in which the phases are
essentially impenetrable walls. Now, this could be a peculiarity of the 
particular selected configuration. In order to make the analysis
quantitative, we shall examine all of our samples and determine the 
strip-crossing probability $P$. This is defined as the probability of
finding a complete path with constant staggered spin across the strip with 
opposite staggered magnetization and we can compute it with the following
algorithm:
\begin{enumerate}
\item For each configuration, compute the Fourier 
transform of the staggered spin field at the smallest nonzero
momentum in each of the three axes ($\phi_x$, $\phi_y$, $\phi_z$).
\item If the system is in a strip configuration, one of
the $\phi$ will be much larger
than the other two. 
Assume this is $\phi_x$, so the strips are perpendicular to the $OX$ axis. 
\item Measure the staggered magnetization $M_\mathrm{s}^x$
 on each of the planes
with constant $x$ and identify the plane $x=x_\mathrm{max}$ 
with largest $|M_\mathrm{s}^x|$. This plane will be at the core
of one the strips.
\item Trace all the clusters that contain at least one spin on
the $x=x_\mathrm{max}$ plane, but severing the links between planes
$x=x_\mathrm{max}$ and $x=x_\mathrm{max}-1$. 
\item If any of the clusters reaches the plane $x=x_\mathrm{max}-1$
there is at least a path through the strip
with opposite magnetization (the previous step has forced us to go
the long way around, so we know we have crossed the strip).
\end{enumerate}

\begin{figure}
\centering
\includegraphics[height=.7\columnwidth,angle=270]{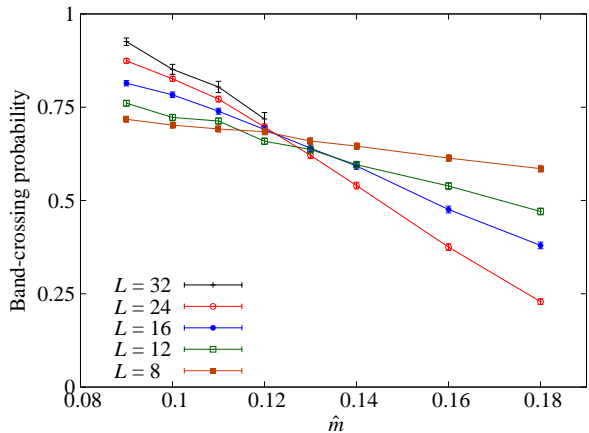}
\caption{(Color online) Probability of completing a path with constant staggered spin across
  the strip with opposite staggered magnetization, see Fig.~\ref{fig:conf},
  for our spin configurations at $\beta=0.625$, $\hat m_\mathrm{s}=0.5$.}
\label{fig:bandas}
\end{figure}

We have plotted the strip-crossing probability $P$ as a function of $\hat m$
in Figure~\ref{fig:bandas}. Notice that $1-P$ behaves as an order parameter.
If we keep increasing $\hat m$, so that we enter the ordered phase, the phases
eventually become proper impenetrable strips, hence $P=0$ for large enough
systems. On the other hand, for low $\hat m$, in the disordered phase, the
strips become increasingly porous, so that $P=1$ in the limit of large
systems. In fact, the inversion of the finite-size evolution at $\hat m\approx
0.12$ signals the onset of the phase transition.

\section{Hard Spheres Crystallization}\label{sec:crystallization}
In this Section we consider a completely different problem, namely
hard-spheres crystallization, as an illustration of the flexibility of the
Tethered Monte Carlo. In particular, this problem has neither quenched
disorder, nor a supporting lattice. Furthermore, the phase transition is
of the first-order, rather than continuous.

Actually, crystallization (or melting) is probably the most familiar example
of a first order transition. Most fluids undergo a transition to a crystalline
solid upon cooling or compression. One might thus be shocked to learn that a
bona-fide Monte Carlo simulation of crystallization, conforming to the
standards taught in good textbooks~\cite{sokal:97}, is plainly impossible
nowadays.

The problem is that there are simply too many local minima of the effective
potential where the simulation can get trapped. Some degree of cheating (you
could also say {\em artistry}) is needed to guide the simulation. 

That the problem is a fairly subtle one is illustrated by the fact that
hard-spheres crystallize in three dimensions~\cite{alder:57,wood:57}, even if
the fluid and the face-centered cubic (FCC) crystals have exactly the same
energy. In fact, hard spheres have become the standard model in the field,
where all new ideas must be tested. Besides their theoretical interest,
hard spheres provide as well an important model for colloidal
suspensions~\cite{pusey:86,pusey:89}.

In fact, workable numerical methods~\cite{vega:08} put by hand the crystal in
the simulation. One may try to achieve equilibrium between the crystal and the
fluid, as in the phase switch Monte Carlo~\cite{wilding:00}, but this is
feasible only for small systems (up to $N=500$ hard
spheres~\cite{errington:04}). One may also compute separately the fluid and
solid free energies. For the fluid, one resorts to thermodynamic integration,
while several methods are available for the crystal
(Wigner-Seitz~\cite{hoover:68}, Einstein crystal~\cite{frenkel:84,polson:00},
Einstein molecule~\cite{vega:07}). The transition point is determined by
imposing to both phases the conditions of equal pressure, temperature and
chemical potential. An alternative is \textit{direct
  coexistence}~\cite{ladd:77,noya:08}, a dynamic, non-equilibrium method that
simulates rather large systems with great accuracy~\cite{zykova-timan:10}.

Our scope here is to illustrate how Tethered Monte Carlo can be used in this
context. Indeed, it has been recently shown that Tethered Monte Carlo allows
equilibrating up to $N=2916$ hard spheres at their phase coexistence
pressure~\cite{fernandez:11}. As explained in the Introduction, constrained
Monte Carlo studies of crystallization kinetics have been carried out
before~\cite{tenwolde:95,chopra:06}, in an umbrella sampling framework.

The remaining part of this section is organized as follows. In
Sect.~\ref{set:hs-model} we recall the hard sphere model. Our order parameters
are defined in Sect.~\ref{sec:hs-order-parameters}. The specific
implementation of the tethered formalism is in
Sect.~\ref{sec:hs-tethered}. Details about the computation of the
phase-coexistence pressure~\cite{fernandez:11} are provided in
Sects.~\ref{sec:p_co-hs} and~\ref{sec:hs-extrema}. The performance of the
Tethered Monte Carlo algorithm is assessed in~\ref{sec:hs-taus}.  Finally, we
comment on the computation of the interfacial free energy in
Sect.~\ref{sec:hs-interfacial}.

\subsection{The hard spheres model}\label{set:hs-model}

We consider a collection of $N$ hard spheres, of diameter $\sigma$.  They are
contained in a cubic simulation box, with periodic boundary conditions. The
system is held at constant pressure $p$ (hence the simulation box may change
its volume, but remaining always cubic). 

Let us introduce the shorthand $\V{R}$ for the set of particle positions,
$\{\V{r}_i\}_{i=1}^N$. The constraint of no overlapping spheres is expressed
with function $H(\V{R})$, which vanishes if any pair of spheres overlaps
($H(\V{R})= 1$ otherwise).\footnote{For a standard model fluid, one would
  replace $H(\V{R})$ by a Boltzmann factor $\exp[-U(\V{R})/(k_\mathrm{B}
    T)]\,.$}

The Gibbs free-energy density, $g(p,T)$, is obtained
from the partition function 
\begin{equation}
Y_{NpT}=\mathrm{e}^{-N \beta g(p,T)}=\frac{p\beta}{ N!
  \Lambda^{3N}}\int_0^\infty\mathrm{d} V \mathrm{e}^{-\beta p V}\int
\mathrm{d} \V{R}\, H(\V{R})\,,
\end{equation}
where  $\Lambda$ is the de Broglie thermal wavelength, while
$\beta=1/(k_\mathrm{B}T)$.

\subsection{The two order parameters}\label{sec:hs-order-parameters}

The standard order parameter for modern crystallization studies is
$Q_6$~\cite{steinhardt:83,duijneveldt:92}, which is the $l=6$ instance
of
\begin{equation} 
Q_l \equiv \left( \frac{4 \pi}{2l +1} \sum_{m = -l}^{l} \left|
Q_{lm} \right|^2 \right)^{1/2}\,,
\end{equation}
where
\begin{equation}
Q_{lm} \equiv \frac{\sum_{i=1
  }^{N}\, q_{lm}(i)}{\sum_{i=1 }^{N} N_b(i)}\,,\quad q_{lm}(i) \equiv
\sum_{j=1}^{N_b(i)} Y_{lm}({\hat{\V{r}}_{ij}})\,.
\end{equation}
In the above expression, $Y_{lm}(\hat{\V{r}}_{ij})$ are the spherical
harmonics, while $\hat{\V{r}}_{ij}$ is the unit vector in the direction
joining particles $i$ and $j$, $\V{r_{ij}}=\V{r}_i-\V{r}_j$.  $N_b(i)$
represents the number of neighbors of the $i$th particle.
\footnote{Two particles $i$ and $j$ are considered neighbors iff
  $r_{ij}<1.5\ \sigma$. This choice ensures that we enclose only the
  first-neighbors shell in the FCC structure, for all the densities of
  interest here~\cite{seoane:12}.} It is important to note that, setting aside
the periodic boundary conditions for our simulation box, $Q_6$ is rotationally
invariant.  $Q_6$ is of order $1/\sqrt{N}$ in a fluid phase, whereas
$Q_6\approx0.574$ or $0.510$ in perfect FCC or BCC crystals, respectively. 
However, defective crystals have lesser values ($Q_6\approx 0.4$ is fairly
common).

Now, if one tries to perform a tethered computation, choosing $Q_6$ as order
parameter, it is soon discovered that it is almost impossible to form an FCC
crystal if the starting particle configuration is disordered. Instead, the
simulation gets stuck in helicoidal crystals, with a similar value of $Q_6$,
which are allowed by the periodic boundary conditions. The crystalline planes
for these helicoidal structures are misaligned with the simulation box.  Upon
reflection, one realizes that the problem lies in the rotational invariance of
$Q_6$. When a crystalline grain starts to nucleate from the fluid, which is
encouraged by the tethered algorithm for large $\hq$, we face the problem that
$Q_6$ conveys no information about the relative orientation of the growing
grain with respect to the simulation box. At some point, the misaligned
crystalline grain is large enough to hit itself through the periodic boundary
conditions, and it needs to contort to get some matching for the crystalline
planes.

The way out is in an order parameter with only cubic symmetry.  Such a
parameter was recently proposed~\cite{angioletti:10}:
\begin{equation}\label{eq:C}
C=\frac{2288}{79}\frac{\sum_{i=1}^{N} \sum_{j=1}^{N_b(i)}c_{\alpha}({\hat{\V{r}}_{ij}})}{\sum_{i=1 }^{N}N_b(i)}-\frac{64}{79} \,,
\end{equation}
where 
\begin{equation}
c_{\alpha}(\hat {\V{r}})=x^4y^4(1-z^4)+x^4z^4(1-y^4)+y^4z^4(1-x^4)\,.
\end{equation}
The expectation value for $C$ in the different phases is the following: $0.0$
in the fluid, $1.0$ in the ideal FCC crystal, perfectly aligned with the
simulation box, and $-0.26$ in the perfectly aligned ideal BCC.  The
difference with the quoted value in Ref.~\cite{angioletti:10} for the perfect
BCC crystal is due to our smaller threshold for neighboring particles. Again,
as it happens for $\q$, we must expect lower $|C|$ values for defective
structures.

In fact, we find that tethered simulations with a single order parameter ($C$)
and large $\hat C$, equilibrate easily. One forms nice crystals even if the
starting particle configuration is disordered, and the Monte Carlo expectation
values turn out to be independent of the starting configuration, as they
should. 

However, not all is well. At some intermediate values of $\hat C$ the
simulation suffers from metastabilities. Heterogeneous states with a slab of
FCC crystal in a fluid matrix are degenerate with helicoidal crystals, that
fill most of the simulation box, but have a small value of $C$ because of
their misalignment. Fortunately, these two types of competing configurations
are clearly differentiated by $\q$, which is large for the helicoidal crystal,
but small for the crystalline slab surrounded by fluid. Therefore, the
combination of the two order parameters labels unambiguously the intermediate states between the fluid and the FCC crystal.

\subsection{Tethered formalism for a hard sphere system}\label{sec:hs-tethered}

Since we wish to constraint simultaneously the two crystal order parameters,
$\hq$ and $\hc$, we use the formalism in Sect.~\ref{sec:several-tethers}. We
choose to couple the demons linearly, which frees us from the constraints $\hq
> Q_6(\V{R})$ and $\hc > C(\V{R})$ imposed by quadratic demons. Note that
ascertaining thermalization is an issue in crystallization studies. It is very
important to compare the outcome of simulations with widely differing starting
configurations. In this respect, the constraints $\hq > Q_6(\V{R})$ and $\hc >
C(\V{R})$ are a major problem, as they prevent us from using the ideal FCC
crystal as starting configuration.

As for the tunable parameter $\alpha$, see
Eq.~\eqref{eq:sum-demons-lineales}, we choose $\alpha=200$.  In this
way, the convolution in the probability distribution functions induced
by the demons does not result into a gross distortion.\footnote{Recall
  that we are convolving in Eq. ~\eqref{eq:convolucion} with a
  Gaussian of width $1/\sqrt{\alpha N}$. The number of spheres that
  can be equilibrated ($N\sim3000$) is rather modest as compared to
  spin models, which suggests enlarging $\alpha$.}

Our tethered statistical weight is
\begin{equation}\label{eq:th-weight-hs}
\omega_N(\V{R},V;p,\hq,\hc)=\frac{N\alpha}{2\pi}
H(\V{R})\ {\mathrm e}^{-\beta pV-N\alpha\, \caja{\paren{\hat{Q_6} -
    Q_6(\V{R})}^2 +\paren{\hat{C} - C(\V{R})}^2}/2}\,.
\end{equation}
As follows from Eq. \eqref{eq:b_lineal}, the gradient of the Helmholtz
effective potential is
\begin{eqnarray}\label{eq:hyomega} 
\grad{\varOmega_N(\hq,\hc,p)}=\paren{\frac{\partial \varOmega_N(\hq,\hc)}{\partial\hq},\frac{\partial \varOmega_N(\hq,\hc)}{\partial\hc}}=\paren{\bigl\langle\hat b_{\q}\bigr\rangle_{\hq,\hc,p},\bigl\langle\hat b_C\bigr\rangle_{\hq,\hc,p}}, \end{eqnarray}
with
\begin{align}\label{eq:B}
 \hat b_{\q}&=\alpha\paren{\hq-\q},& 
\hat b_{C}&=\alpha\paren{\hc-C}.
\end{align}

The relationship between the effective potential, $\Omega_N(\hq,\hc,p)$, and
the Gibbs free-energy density is straightforward:
\begin{eqnarray} Y_{NpT}=e^{-N
  g_N(p,T)}=\int\mathrm{d}\hq\ \mathrm{d}\hc\ e^{-N\varOmega_N(\hq,\hc,p)}\,.\end{eqnarray}
Furthermore, a saddle-point approximation, see
Sect.~\ref{sec:ensemble-equivalence}, shows that
\begin{eqnarray} \label{eq:geqomega} g_{N}(p,T)=\varOmega_N(\hq^*,\hc^*,p)+O(1/N)\,,
\end{eqnarray} 
where $(\hq^*,\hc^*,p)$ corresponds to the $p$-dependent absolute minimum of
$\varOmega_N(\hq,\hc,p)$, regarded as a function of $\hq$ and $\hc$. However,
it is important to keep in mind that near the phase transition two local
minima (namely the fluid and the FCC crystal) become exactly degenerate.

\subsection{The coexistence pressure: computing differences in the effective potential}\label{sec:p_co-hs}

\begin{figure}
\includegraphics[angle=270,width=\columnwidth,trim=0 0 0 0]{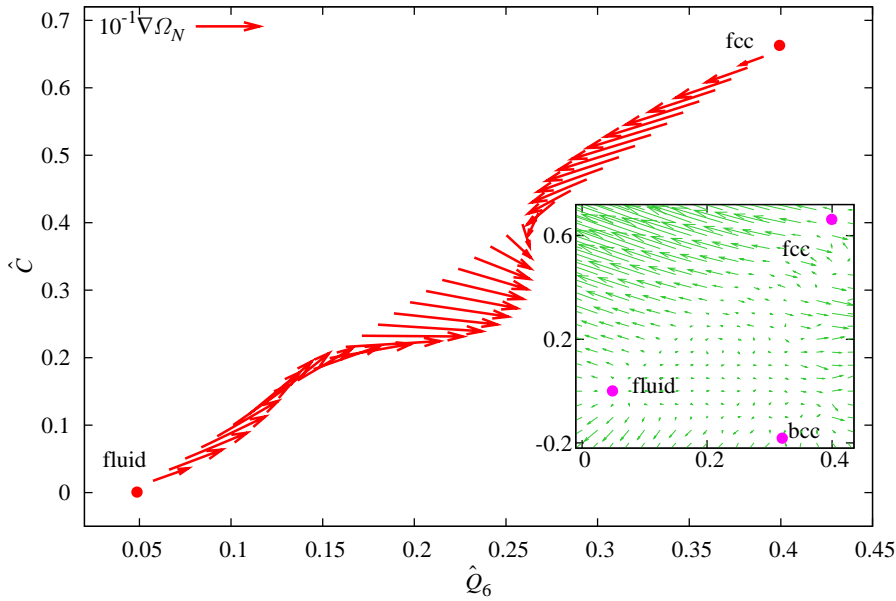}
\caption{(Color online) $\grad\varOmega_{N=256}$, as computed from Eq.~\eqref{eq:B} for
  $N=256$ hard spheres, along the straight path that joins the fluid and the
  FCC minima of the effective potential. The simulation pressure is the
  phase-coexistence one. To improve visibility, we have divided
  $\grad\varOmega_{N=256}$ by a factor of 10. {\bf Inset:} map of the gradient field
  $\grad\varOmega_{N=256}$, including the points corresponding to the fluid,
  FCC and BCC potential minima. For the sake of visibility we have divided the
  gradient by a factor $\alpha=200$ (mind the different normalization as
  compared with main panel). }
\label{fig:grid}
\end{figure}

The coexistence pressure, $p_\mathrm{co}$, follows from the difference in
effective potential between the pressure-dependent coordinates of
the coexisting pure phases:
\begin{equation}\label{eq:delta-omega-hs}
\Delta\Omega_N(p)=\Omega_N\bigl(\hq^\mathrm{FCC}(p),\hc^\mathrm{FCC}(p),p\bigr)-
\Omega_N\bigl(\hq^\mathrm{fluid}(p),\hc^\mathrm{fluid}(p),p\bigr)\,.
\end{equation}
The scope of the game is finding the coexistence pressure,
$p_\mathrm{co}^{N}$, such that $\Delta\Omega_N(p_\mathrm{co}^{N})=0$.  Indeed,
the saddle-point condition \eqref{eq:geqomega}, tells us that, at
$p_\mathrm{co}^{N}$, the chemical potential for the two phases coincides.

Now, the computation of $\Delta\Omega_N(p)$ is strictly analogous to that of
section~\ref{sec:daff-th}. We obtain the difference in the effective potential
as a line integral of the conservative field $\hat{\boldsymbol{B}}$. The most
convenient path is a straight line, see Fig. \ref{fig:grid}. As in
Sect.~\ref{sec:daff-th}, we divide the path in a mesh, perform independent
simulations at each point of the grid, and compute $\Delta\Omega_N(p)$ from a
numerical integration of the gradient, Eq.~(\ref{eq:hyomega}), projected over
the path.

Let us stress only the main difference with the computation in
Sect.~\ref{sec:daff-th}:
\begin{itemize}
\item Here, we wish to compute $\Delta\Omega_N(p)$ as a function of
  pressure. Instead, in section~\ref{sec:daff-th} we were restricted to a
  single value of the control parameter (temperature in that case).

\item The disorder-averaging procedure introduced in
  Sect.~\ref{sec:DAFF-disorder-average} dispensed us from a careful search
of local minima. Instead, locating with very high accuracy the extremal points
  $\bigl(\hq^\mathrm{FCC}(p),\hc^\mathrm{FCC}(p)\bigr)$ and
  $\bigl(\hq^\mathrm{fluid}(p),\hc^\mathrm{fluid}(p)\bigr)$ is a must
  here, see Sect.~\ref{sec:hs-extrema}. 
  
\item The choice of a straight line as integration path, see
  Fig. \ref{fig:grid}, is not only dictated by simplicity. It is also a matter
  of computational convenience. Thermalization is easier to achieve over
  it. Furthermore, as the gradient map in the inset of Fig. \ref{fig:grid}
  shows, the gradient takes its smallest absolute value over this path.
\end{itemize}

In order to compute $p_\mathrm{co}^N$, let us neglect the pressure dependence
of the end points for the integration path in Fig.~\ref{fig:grid}.  One may
easily correct for end-points displacements, as explained in
Sect.~\ref{sec:hs-extrema}, which induces a correction in $p_\mathrm{co}^N$
negligible with respect to our statistical errors.

Under the above simplifying assumption, we only need the pressure dependence
of the gradient field, $\hat{\boldsymbol{B}}$, over the straight path in
Fig.~\ref{fig:grid}. Using histogram
reweighting~\cite{falcioni:82,ferrenberg:88}, we extrapolate our numerical
results at pressure $p$ to a neighboring $p+\delta p$:
\begin{equation}\label{eq:reweighting-hs}
\langle \hat{\boldsymbol{B}} \rangle_{\hq,\hc,p+\delta p}=\frac{\langle
  \hat{\boldsymbol{B}} \mathrm{e}^{-\delta p V}\rangle_{\hq,\hc,p}}
{\langle\mathrm{e}^{-\delta p V}\rangle_{\hq,\hc,p}}.
\end{equation}
A standard argument tells us that the maximum safe extrapolation, $\delta p$,
is determined by the probability distribution function (pdf) for the specific-volume,
$v=V/N$,
\begin{equation}
\delta p^\mathrm{maximum}\approx \frac{1}{\sqrt{N\chi_p}}\,,\quad
\chi_p= \left.\frac{\partial \langle v\rangle}{\partial p}\right|_{(\hq,\hc,p)}=N\bigl[\langle v^2\rangle_{\hq,\hc,p} -\langle v\rangle^2_{\hq,\hc,p}\bigr]\,.
\end{equation}
Hence, it is crucial that the pdf for $v$ be unimodal (i.e. single-peaked),
and with an $N$-independent $\chi_p$, for all point along the integration
path. In other words, it is important that the integration path  be free
of metastabilities. Since this condition holds~\cite{seoane:12}, it is
straightforward to compute $\Delta\Omega_N(p+\delta p)$ from simulations at
$p$.

Following these guidelines, $p_\mathrm{co}^N$ was computed in
Ref.~\cite{fernandez:11} for $108\leq N\leq 2916$.  The large $N$
extrapolation was
\begin{center}
$p_\mathrm{co}^{\infty}=11.5727(10)$~\cite{fernandez:11}\,.
\end{center}
The best previous equilibrium estimate seems to be the rather crude
$p_\mathrm{co}^\infty=11.50(9)$~\cite{wilding:00}, obtained using phase-switch
Monte Carlo. In fact, the only previous method accurate enough to provide a
meaningful comparison is the non-equilibrium direct-coexistence:
$p_\mathrm{co}^{\infty}=11.576(6)$~\cite{zykova-timan:10}. Note, however, that
in order to achieve such a small error (but still six times larger than the
tethered error), systems with up to $N=1.6\times 10^5$ particles were
simulated~\cite{zykova-timan:10}.

\subsection{Calculation of the extremal points and corrections}\label{sec:hs-extrema}

We need to locate the two extremal points in the straight path in
Fig.~\ref{fig:grid}, which correspond to the fluid or to the FCC crystal. The
two points are local minima of $\Omega_N$, regarded as a function of $\hq$ and
$\hc$ but at fixed pressure.  Our procedure has been as follows.

We first obtain a crude estimate from standard simulations in the $NpT$
ensemble (without any constrain in the crystal parameters). Note that the
autocorrelation time for such simulations is unknown, but larger than any
simulation performed to date. Hence, these standard simulations get stuck at
the local minimum of $\Omega_N$ which is most similar to their starting
configuration. Starting the simulation either from an ideal gas, or from a
perfect FCC crystal, we approach the pure-phases we are interested in. The
Monte Carlo average of $\q(\V{R})$ and $C(\V{R})$ provides our first guess.

To refine the search of either of the two local minima $(\hq^*,\hc^*)$, we
note that, up to terms of third order in $\hq -\hq^*$ or $\hc -\hc^*$,
\begin{equation}\label{eq:gaussian-aprox-hs}
\Omega_N(\hq,\hc)=\Omega_N^*+
\frac{A_{QQ}}{2}(\hq-\hq^*)^2+A_{QC}(\hq-\hq^*)(\hc-\hc^*)+\frac{A_{CC}}{2}(\hc-\hc^*)^2\,.
\end{equation}
The shorthand $\Omega_N^*$ stands for $\Omega_N(\hq^*,\hc^*)$.  Incidentally,
Eq.~\eqref{eq:gaussian-aprox-hs} tells us that the computation in
Sect.~\ref{sec:p_co-hs} is intrinsically stable. An error of order $\epsilon$
in the location of $(\hq^*,\hc^*)$ will result in an error of order
$\epsilon^2$ in the coexistence pressure.

Yet, the tethered computation does not give us access to $\Omega_N$, but to its
gradient:
\begin{equation}\label{eq:gaussian-gradient-hs}
\grad \Omega_N(\hq,\hc)= \bigl(A_{QQ} (\hq-\hq^*)+A_{QC}(\hc-\hc^*), A_{CC}(\hc-\hc^*)+A_{QC}(\hq-\hq^*)\bigr)\,.
\end{equation}
Eq.~\eqref{eq:gaussian-gradient-hs} holds up to corrections quadratic in $\hq
-\hq^*$ or $\hc -\hc^*$. We thus compute the expectation value of the field
$\hat{\boldsymbol{B}}$, in a grid of nine points $(\hq,\hc)$ that surround our
first guess for $(\hq^*,\hc^*)$, and fit the results to
Eq.~\eqref{eq:gaussian-gradient-hs}. We iterate this procedure until an
accuracy $\sim 10^{-6}$ in both coordinates $(\hq^*,\hc^*)$ is
reached~\cite{seoane:12}.  

Actually, Eq.~\eqref{eq:reweighting-hs}, shows how one extrapolates the
expectation values for the gradient field from the simulated pressure, $p$ to
a nearby $p+\delta p$. The corresponding fit to
Eq.~\eqref{eq:gaussian-gradient-hs} provides the new coordinates
$\bigl(\hq^*(p+\delta p),\hc^*(p+\delta p)\bigr)$.

At this point, one could worry because the integration path in
Fig.~\ref{fig:grid} is no longer appropriate at pressure $p+\delta p$. In
fact, the extremal points in the integration path are pressure-dependent.
However, some reflection shows that this is not a real problem. In fact,
\begin{equation}
\Delta\Omega_N(p+\delta p)= \Delta^\mathrm{FCC}(p,p+\delta p)+
\Delta^\mathrm{path}(p,p+\delta p)-\Delta^\mathrm{fluid}(p,p+\delta p)\,.
\end{equation}
The different pieces are
\begin{eqnarray}
\Delta^\mathrm{FCC}(p,p+\delta p)&=&
\Omega_N\bigl(\hq^\mathrm{FCC}(p+\delta p),\hc^\mathrm{FCC}(p+\delta
p);\,p+\delta p\bigr)-\\\nonumber
&-&\Omega_N\bigl(\hq^\mathrm{FCC}(p),\hc^\mathrm{FCC}(p);\,p+\delta
p\bigr)\,,
\end{eqnarray}
the correction due to the shift of order $\delta p$ in the coordinates of the
FCC minimum,
\begin{eqnarray}
\Delta^\mathrm{path}(p,p+\delta p)&=&
\Omega_N\bigl(\hq^\mathrm{FCC}(p),\hc^\mathrm{FCC}(p);\,p+\delta p\bigr)-\\\nonumber
&-&\Omega_N\bigl(\hq^\mathrm{fluid}(p),\hc^\mathrm{fluid}(p);\,p+\delta
p\bigr)\,,
\end{eqnarray}
the line-integral sketched in Fig.~\ref{fig:grid} as computed at pressure
$p+\delta p$, and
\begin{eqnarray}
\Delta^\mathrm{fluid}(p,p+\delta p)&=&
\Omega_N\bigl(\hq^\mathrm{fluid}(p+\delta p),\hc^\mathrm{fluid}(p+\delta
p);\,p+\delta p\bigr)-\\\nonumber
&-&\Omega_N\bigl(\hq^\mathrm{fluid}(p),\hc^\mathrm{fluid}(p);\,p+\delta
p\bigr)\,,
\end{eqnarray}
the correction due to the shift in the coordinates of the fluid minimum.

Now, one expects that the pressure-induced changes in the minima coordinates
as well as on the coefficients $A_{QQ}$, $A_{Q,C}$ and $A_{CC}$ will of order
$\delta p$. Hence, Eq.~\ref{eq:gaussian-aprox-hs} implies that both
$\Delta^\mathrm{FCC}(p,p+\delta p)$ and $\Delta^\mathrm{fluid}(p,p+\delta p)$
are of order $(\delta p)^2$. This is the rationale behind the simplifying
assumption made in Sect.~\ref{sec:p_co-hs}.

At any rate, $\Delta^\mathrm{FCC}(p,p+\delta p)$ and
$\Delta^\mathrm{fluid}(p,p+\delta p)$ can be numerically computed from
Eq.~\ref{eq:gaussian-aprox-hs}. For all values of $N$ simulated in
Ref.~\cite{fernandez:11}, their combined effect on the determination of the
coexistence pressure turns out to be smaller than 1\% of the statistical
error bars~\cite{seoane:12}.

\subsection{Algorithmic performance}\label{sec:hs-taus}

\begin{figure}
\includegraphics[angle=270,width=0.95\columnwidth,trim=0 0 0 0]{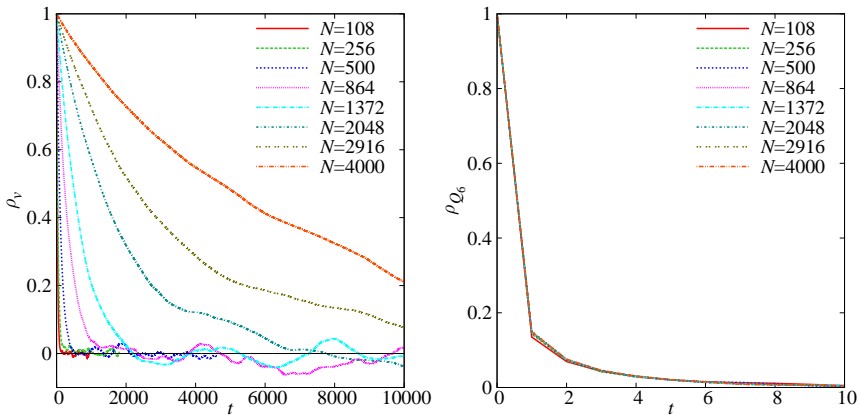}
\caption{(Color online) Normalized time autocorrelation function,
  Eq.~\ref{eq:C-O-t}, for the specific volume (left) and the crystal
  order parameter (right), as computed for a system of $N$ hard
  spheres, in the fluid minimum of the effective potential (labeled
  $S=0$). Time is measured in units of EMCS (see text). Mind the
  different time scale for the left and right panels. Each value of
  $N$ was simulated very close to (but not precisely at) its
  phase-coexistence pressure $p_\mathrm{co}^N$~\cite{fernandez:11}.}
\label{fig:autocorr}
\end{figure}

The simulation of the weight in Eq.~\eqref{eq:th-weight-hs} requires two types
of moves: single particle displacements, as well as changes in the volume of
the simulation box. We shall use the short hand Elementary Monte Carlo Step
(EMCS) to the combination of $N$ consecutive single-particle displacements
attempts,\footnote{We pick at random a particle-index, say $i$, and try
  $\V{r}_i\rightarrow \V{r}_i+\V{\delta}$ with $\V{\delta}$ chosen with
  uniform probability within the sphere of radius $\Delta$. We tune $\Delta$
  to keep the acceptance above $30\%$.} followed by a change attempt in the
simulation box volume.

The standard tools to asses the performance of a Monte Carlo
algorithm~\cite{sokal:97} are briefly recalled in the
Appendix~\ref{sec:thermalization}. 

Recall that we will be discussing independent simulations along the path in
Fig.~\ref{fig:grid}. We thus will be labeling them by means of a coordinate
$S$, such that $S=0$ corresponds to the fluid pure phase, while $S=1$ refers
to the FCC minimum.

One should like to consider the time autocorrelation functions for the
components of the gradient field, $\hat b_{\q}$ and $\hat b_C$. Yet,
Eq.~\eqref{eq:B} tells us that these correlation functions are identical to
those of $Q_6(\V{R})$ and $C(\V{R})$. Eq.~\eqref{eq:reweighting-hs} suggests as
well that the time autocorrelation function for the specific volume $v$ is of
interest. An example of these autocorrelation functions is shown in
Fig.~\ref{fig:autocorr}, for the $S=0$ point. We note that $v$ plays the role
of the algorithmic slow mode, with a strong $N$ dependence. On the other hand,
the autocorrelation function for $Q_6$ decreases very fast, and it is barely
$N$-dependent. The autocorrelation function for $C$ is qualitatively
identical to that of $Q_6$, and will thus be skipped.

\begin{figure}
\centering
\includegraphics[angle=270,width=0.85\columnwidth,trim=0 0 0 0]{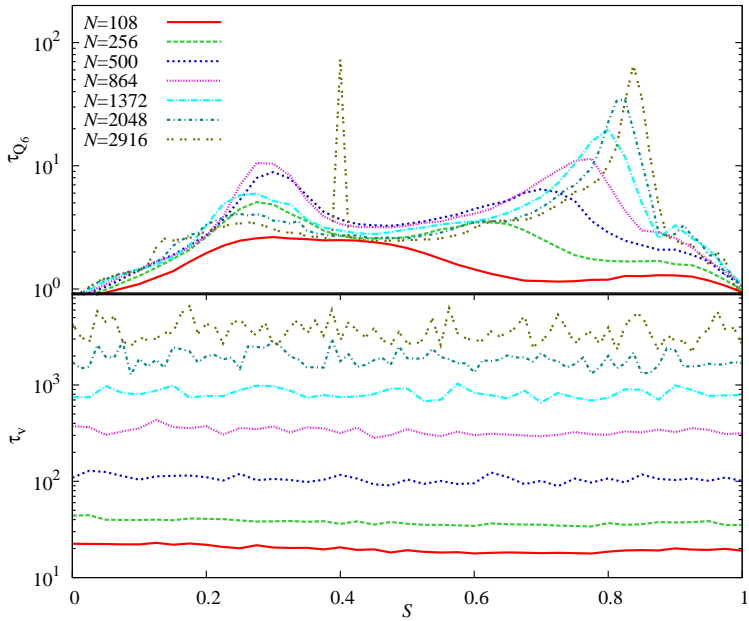}
\caption{(Color online) Integrated autocorrelation times, defined in
  Eq.~\ref{eq:app-tau-int}, for $\q$ (top) and $v$ (bottom), as a function of
  $S$ (the linear coordinate that labels the integration path in
  Fig.~\ref{fig:grid}, where $S=0$ stands for the fluid minimum and $S=1$
  represents the homogeneous FCC phase). Time is measured in units of
  EMCS. Results for all values of $N$ equilibrated in
  Ref.~\cite{fernandez:11}.}
\label{fig:tiempos}
\end{figure}

The analysis is made quantitative by considering the integrated
autocorrelation times, see Fig.~\ref{fig:tiempos}.  We notice that the
dynamics of $v$ is considerable slower than that of $\q$, and featureless as a
function of $S$. Data for the specific volume scales as $\tau_v\sim N^{5/3}$
(quite worse than standard critical slowing down in three dimensions,
$\tau\sim N^{2/3}$, yet much better than exponential dynamic
slowing-down). There is a clear anomaly in the behavior of $\tau$ for a
single simulation point in $N=2916$. We briefly comment on this below.

Using these tools, it was ensured in Ref.~\cite{fernandez:11} that all
simulations were, at least, $50\tau$ long. Besides, all simulations were
performed twice, with different starting configurations (either an ideal FCC
crystal, or an ideal gas). Compatibility between the two sets of
investigations was systematically checked~\cite{seoane:12}.

The anomaly at $S=0.4$ for $N=2916$ is due to the emergence of metastability.
At this value of $S$ we expected to find a spatially segregated state (a slab
of FCC crystal in a liquid matrix). This state appeared indeed, but the
simulation tunnels back and forth from it to an helicoidal crystal (in close
analogy with the Monte Carlo history shown in the bottom-left panel of
Fig.~\ref{fig:metaestabilidad}). We actually performed extra simulations in
this point, in order to determine the relative weight of each of the two
metastable states (their effect was carefully considered in final
estimates~\cite{seoane:12}).  

These helicoidal crystals appear much more often for $N=4000$ and intermediate
$S$. Nevertheless, selecting carefully the starting particle configuration for
the simulation at each $S$, one may obtain a gradient field with a smooth
$S$-dependency. However, it is clear that these $N=4000$ results, although
plausible, cannot be regarded as well equilibrated~\cite{fernandez:11}.

\subsection{Interfacial free-energy}\label{sec:hs-interfacial}

\begin{figure}
\centering
\includegraphics[angle=270,width=0.75\columnwidth,trim=0 0 0 0]{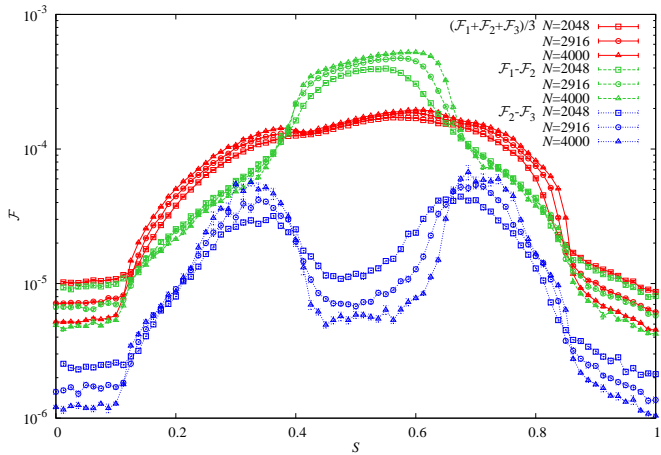}
\caption{(Color online) For systems of $N$ hard spheres at their phase-coexistence pressure,
  we show, as a function of the line parameter $S$, different linear
  combinations of the particle-density fluctuations, Eq.~\eqref{eq:hs-F-def},
  computed for the minimal wavevectors allowed by periodic boundary conditions
  and ordered in such a way that ${\cal F}_1>{\cal F}_2>{\cal F}_3$. For the
  phase-separated states all three ${\cal F}$ are of order $1$ (order $1/N$
  for homogeneous systems). The slab phase is the only one with ${\cal
    F}_1-{\cal F}_2$ of order one. The cylinder phase is identified by ${\cal
    F}_2-{\cal F}_3$ of order one.}
\label{fig:F}
\end{figure}

The interfacial free energy is the free-energy cost per unit area of a
liquid-to-crystal interface. Its computation has been rather difficult for
hard spheres, different authors finding mutually incompatible
results~\cite{davidchack:00,mu:05,davidchack:10}.

The tethered formalism allows as well to compute the interfacial
free energy~\cite{fernandez:11}. One should identify a point along the
straight path in Fig~\ref{fig:grid} where a  slab of fluid surrounded by an FCC
crystal is formed. Such configurations have two phase-separating surfaces of
linear dimensions comparable with those of the simulation box. The effective
potential difference between such a heterogeneous state and either of the two
pure phases provides an estimate of the interfacial free energy.  It turns out
that a well formed surface appears only for $N\geq 2048$~\cite{fernandez:11}.
Let us see how this can be ascertained.

The physical situation is as follows. When we go from the liquid to the solid,
Fig.~\ref{fig:grid}, the homogeneous fluid becomes unstable at a value of the
linear coordinate $S\propto N^{-1/(D+1)}$, which means that a macroscopic
droplet of crystal forms. This has been established for all types of
first-order phase
transitions~\cite{biskup:02,binder:03,macdowell:04,nussbaumer:06}, and
explicitly verified for crystallization~\cite{fernandez:11}. As $S$ grows the
mass of the crystal droplet increases, which costs surface energy. At a
certain point, the periodic boundary conditions allow reducing the surface
energy by turning the crystal droplet onto a crystal cylinder. At still larger
$S$, the cylinder becomes a slab. Of course another three analogous
geometrical transitions arise when $S$ keeps increasing as we approach the FCC
minimum. All six geometric transitions appeared in large enough hard-spheres
systems~\cite{fernandez:11}. We are interested in identifying systems large
enough to form a slab of crystal surrounded by fluid.

To follow these geometric transitions we consider the particle-density
fluctuations quantified through
\begin{equation}\label{eq:hs-F-def}
{\cal F}(\V{q})=\frac{1}{N^2} \biggl|\sum_{i=1}^N e^{\mathrm{i} \V{q}\cdot\V{r}_i}\biggr|^2\,,
\end{equation}
As we are interested in the largest wavelength, we consider the smallest
$\V{q}$ allowed by periodic boundary conditions, $\Vert \V{q}\Vert= 2\pi/L$,
where $L$ is the linear size of the simulation box. There are three such
minimal wavevectors in a cubic box. Given a particle configuration, ${\cal
  F}_1$ is the maximum over the three directions, ${\cal F}_3$ is the minimum,
and ${\cal F}_2$ is the intermediate one. As the droplet, cylinder and slab
geometries have different symmetries the natural order parameters are
\begin{itemize}
\item Whenever the system is phase separated,
  $(\mathcal{F}_1+\mathcal{F}_2+\mathcal{F}_3)/3$ is of order 1, (order $1/N$
  otherwise).
\item For a cylinder, two of the ${\cal F}$'s are of order 1, while the ${\cal
  F}$ along the cylinder axis is small. Hence, $\mathcal{F}_2-\mathcal{F}_3$
  is of order 1 in the cylinder phase, but it vanishes (for large $N$) both in
  the droplet and the slab phase.
\item For a slab the only ${\cal F}$ of order 1 is that transversal to
  it. Hence $\mathcal{F}_1-\mathcal{F}_2$ is of order 1 for a slab, but not
  for the cylinder nor the droplet.
\end{itemize}
All these behaviors are identified in Fig.~\ref{fig:F}. We thus conclude
that $N\geq 2048$ is sufficient to attempt a computation of the interfacial free-energy.

To improve the numerical stability, we consider the potential difference
between two points, $S^*\approx 0.5$ and $S=1$. For both $S$ values, the
gradient of $\Omega_N$ is normal to the straight in Fig~\ref{fig:grid}
(actually $S=1$ is a local minimum of $\Omega_N$, while $S^*$ is a local
maximum if we restrict ourselves to the straight).  Hence, the interfacial
free-energy for the $(100)$ crystal direction, $\gamma_{100}$, follows
from~\cite{fernandez:11}
\begin{equation}\label{eq:hs-gamma}
\frac{\gamma_{100}}{k_\mathrm{B}T}= N^{1/3}
\frac{\paren{\Omega_\text{FCC}-\Omega_{S^*}}}{2\mean{v}_{S^*}^{2/3}}  \,.
\end{equation}

\begin{figure}
\centering
\includegraphics[angle=270,width=0.75\columnwidth,trim=0 0 0 0]{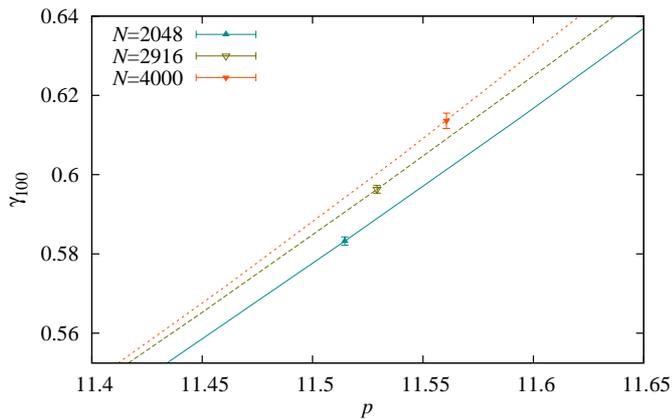}
\caption{(Color online) Interfacial free-energy for the $(100)$ lattice crystalline direction
  $\gamma_{100}$ as a function of pressure, for a system of $N$ hard-spheres,
  for $N=2048,2916$ $N=4000$. We estimated $\gamma_{100}$ from
  Eq.~\ref{eq:hs-gamma}}
\label{fig:sigma_p}
\end{figure}

A peculiarity of the tethered approach is that one may control the dependence
of the estimate of $\gamma_{100}$ on the actual estimate used for the
coexistence pressure. One simply computes $\gamma_{100}$ as a function of
pressure, using Eq.~\ref{eq:hs-gamma}, as it is shown in
Fig.~\ref{fig:sigma_p}. It turns out that the slope of the curve is of order
$0.4$, hence an error of order $\epsilon$ in the determination of
$p^\infty_\mathrm{co}$ results in an error of order $\sim 0.4 \epsilon$ in
$\gamma_{100}$. To our knowledge, such effects have not been taken into
account in previous computations~\cite{davidchack:00,mu:05,davidchack:10}.

\section{Conclusions}\label{sec:conclusions}

In this work, we have explained Tethered Monte Carlo~\cite{fernandez:09}, a
systematic and very general methodology to overcome free-energy barriers in
Monte Carlo simulations without a random-walk in the {\em reaction-coordinate}
space. An admittedly weak point is that a deep physical insight is needed in
order to identify a suitable reaction coordinate. Nevertheless, it is
worth mentioning that this problem may be bypassed using real
replicas~\cite{cammarotta:10}.

The power of the method is illustrated in two famous problems of modern
theoretical physics, the diluted antiferromagnet in an external field (DAFF),
the physical realization of the random field Ising model, and the
crystallization of hard spheres. In both cases, the tethered approach allows
to go beyond the state of the art.

Let us stress that Tethered Monte Carlo is not limited to reproducing
canonical mean values, but it is also suitable to address different physical
questions that one would never ask in a canonical setting due to overwhelming
technical difficulties. Our study of the pseudo phase coexistence in the DAFF,
see Sect.~\ref{sec:DAFF-geometry} is one example. Furthermore, for disordered
systems, tethered Monte Carlo provides a redefinition of the quenched
disorder, less vulnerable to (trivial) violations of
self-averaging~\cite{fernandez:08,fernandez:11b}.

We remark as well that, from the algorithmic point of view, we have been
rather unsophisticated. We used local Metropolis moves that (in the case of
the DAFF) were complemented with parallel tempering. Clearly enough there is
much room for improvements on this side (see e.g.~\cite{martin-mayor:09} for
an implementation of a cluster algorithm).

We also note that, in our implementation, both for the DAFF and for hard
spheres there is a maximum system size beyond which one cannot equilibrate
within reasonable computer time. Clearly enough new free-energy barriers
appear, which are not adequately described by the chosen reaction
coordinates. These could be treated by extending the number (or the type) of
tethered quantities.

\begin{acknowledgements}

Many of the ideas discussed here evolved from work done in collaboration with
L.A. Fernandez, A. Gordillo-Guerrero, J.J. Ruiz-Lorenzo and P. Verrocchio.  We
are also thankful to L.G. MacDowell, C. Vega and G. Parisi for discussions.

Simulations were performed at BIFI ({\em Terminus}), at the Red Espa\~nola de
Supercomputaci\'on ({\em Marenostrum}), and at {\em Piregrid} and {\em
  Ibergrid}. We thank these institutions for the computer resources, technical
expertise and assistance provided.

We acknowledge partial financial support from MICINN, Spain, through research
contract no. FIS2009-12648-C03 and from UCM-Banco de Santander (contract
no. GR32/10-A/910383). B.S. and
D.Y. were supported by the FPU program.

\end{acknowledgements}

\appendix
\section{Assessing thermalization in Monte Carlo simulations}\label{sec:thermalization}
This appendix is intended as a brief and handy reference on thermalization 
in Monte Carlo simulations, providing some definitions that 
are used throughout the paper (see~\cite{sokal:97} for a detailed
discussion). Most of the definitions are standard, 
but subsection~\ref{sec:thermalization-PT} describes a powerful
thermalization criterion that 
may be new even for the Monte Carlo specialist.

As a general rule, the thermalization of a Monte Carlo simulation should be 
assessed through the temporal autocorrelation functions~\cite{sokal:97}.
If $O(t)$ is the measured value of the observable $O$ at time
$t$ during the simulation, then
\begin{equation}\label{eq:C-O-t}
C_O(t,\hat m) =\bigl\langle[O(0)-\langle O\rangle_{\hat m}][O(t)-\langle O\rangle_{\hat m}]\bigr\rangle_{\hat m},   ,\qquad
\rho_O(t,\hat m) = \frac{C_O(t,\hat m)}{C_O(0,\hat m)}\ .
\end{equation}
From this function we obtain the integrated and exponential 
autocorrelation times
\begin{align}
\tau_{\text{int},O}(\hat m) &= \frac12 + \sum_{t=1}^\infty \rho_O(t,\hat m),\label{eq:app-tau-int}\\
\tau_{\text{exp},O}(\hat m) &= \lim_{t\to\infty} \sup \frac{t}{-\log|\rho_O(t,\hat m)|},\\
\tau_\text{exp}(\hat m) &= \sup_{O} \tau_{\text{exp},O} (\hat m).
\end{align}
In general, the autocorrelation function can be expressed as a sum of
exponentials\footnote{% Strictly speaking the decomposition in
  Eq.~\eqref{eq:rho-exponentials} holds only for an aperiodic dynamics
  that fullfils detailed balance.  Violation of either of these two
  condicions will result in the appearance of complex
  eigentimes$\tau_i$. If the dynamics is aperiodic, the corresponding
  eigenvalue of the generator of the Markov process,
  $\ee^{-1/\tau_i}$, will be smaller than one in absolute value. Hence
  for these modes, the correlations could take the form of a damped
  oscillation. A well known example is the autocorrelation function
  for the magnetization in the Wolff single-cluster dynamics for the
  Ising model (see, e.g., Fig. 3--5 in~\cite{amit:05}). Indeed,
  cluster algorithms verify balance, but not detailed balance, because
  the corresponding generator of the Markov chain is formed by
  \emph{alternating} steps of two detailed-balance fullfiling
  dynamics, namely bond-updating and spin-updating. In the case of a
  detailed-balance dynamics, the eigenvalues $\ee^{-1/\tau_i}$ are
  real but they could still be negative.  Having said so, in most
  applications (and certainly in the ones considered here), oscilating
  behavior is not observed within statistical errors.}
\begin{equation}\label{eq:rho-exponentials}
\rho_O(t,\hat m) = \sum_i A_i \ee^{-t/\tau_i}.
\end{equation}
The exponential time, then, is the largest of the $\tau_i$ and characterizes
the time needed to equilibrate a certain observable (or the whole system).
The integrated time indicates the minimum time difference so that
two measurements of some observable $O$ can be considered 
independent (i.e., uncorrelated). Notice that if the decomposition \eqref{eq:rho-exponentials}
contains a single exponential, $\tau_{\text{exp},O} = \tau_{\text{int},O}$,
This 
is a useful observation because the exponential time is harder to measure
accurately than the integrated one.

\subsection{Thermalization in parallel tempering simulations}\label{sec:thermalization-PT}
The computation of the $\tau_O$
requires a much longer simulation time than what
is needed to thermalize the system. This is not a problem for ordered
systems, since extending the simulation reduces the statistical errors.
In disordered systems, however,
the errors are dominated by sample-to-sample fluctuations, so 
an efficient CPU time allocation should run each sample for a time
long enough to thermalize it, but no longer. Therefore, the computation
of autocorrelation times in these cases has traditionally been abandoned 
as impractical and thermalization is typically assessed by less 
rigorous methods such as the time evolution of disorder-averaged 
observables. 

The use of parallel tempering, however, provides a way to 
compute the autocorrelation times in relatively short simulations.
The key is analyzing quantitatively the temperature random walk of
each participating configuration~\cite{fernandez:09b}. Reference~\cite{janus:10}
includes a detailed discussion of this method, here we shall only give
a brief sketch of it.

Let us consider a parallel tempering simulation  
with $N_T$ participating replicas of the system.  We define $f_i(t)$ as 
the function indicating the temperature index of replica $i$ at time $t$.
By `temperature index' we mean the temperature mapped onto 
$\{1,2,\ldots,N_T\}$, where $1$ corresponds to the lowest temperature
and $N_T$ to the highest. We can now compute 
the autocorrelation function~\eqref{eq:C-O-t} of each $f_i$ 
(notice that $\langle f_i\rangle = (N_T+1)/2$). The $N_T$ correlation
functions thus obtained can be averaged, so they collectively 
have the precision of a much longer simulation.

\end{document}